\documentclass[%
 aip,
 amsmath,amssymb,
 reprint,%
]{revtex4-2}

\usepackage{graphicx}
\usepackage{dcolumn}
\usepackage{bm}

\usepackage[utf8]{inputenc}
\usepackage[dvipsnames]{xcolor}
\usepackage{soul}
\RequirePackage[normalem]{ulem}
\usepackage{hyperref}
\usepackage[nameinlink]{cleveref}
\usepackage{siunitx}
\usepackage{array}
\usepackage{upgreek}
\usepackage{tabularx}

\Crefname{equation}{Eq.}{Eqs.}
\Crefname{figure}{Fig.}{Figs.}
\Crefname{table}{Tab.}{Tabs.}

\makeatletter
\def\@email#1#2{%
 \endgroup
 \patchcmd{\titleblock@produce}
  {\frontmatter@RRAPformat}
  {\frontmatter@RRAPformat{\produce@RRAP{*#1\href{mailto:#2}{#2}}}\frontmatter@RRAPformat}
  {}{}
}%
\makeatother

\def\doctitle{Photorefractive tuning seeded by third-harmonic light in a diamond photonic crystal cavity}

\def\authorOne{Joe Itoi}
\def\authorTwo{Elham Zohari}
\def\authorThree{Nicholas J. Sorensen}
\def\authorFour{Waleed El-Sayed}
\def\authorFive{Joseph E. Losby}
\def\authorSix{Gustavo O. Luiz}
\def\authorSeven{Sigurd Fl{\aa}gan}
\def\authorEight{Paul E. Barclay}
\def\authorNine{Sean McNaney}
\def\addressOne{
    Institute for Quantum Science and Technology, University of Calgary, Calgary, AB, T2N 1N4, Canada
    }
\def\addressTwo{
    Department of Physics, University of Alberta, Edmonton, Alberta T6G 2E1, Canada
    }
 \def\addressThree{
National Research Council Canada, Quantum and Nanotechnology Research Centre, Edmonton, Alberta T6G 2M9, Canada
}
\def\addressFour{
nanoFAB Centre, University of Alberta, Edmonton, Alberta T6G 2V4, Canada
}

\def\emailContactOne{sigurd.flagan@ucalgary.ca}
\def\emailContact{pbarclay@ucalgary.ca}

\def\abstractText{
Single-crystal diamond nanocavities have tremendous potential for use in quantum and nonlinear optical technologies. 
The ability to precisely control their resonant frequencies is essential for many applications, and $\textit{in situ}$ tuning is particularly desirable.  
In this work, we demonstrate deterministic resonance tuning of a diamond nanocavity. 
We observed a photorefractive effect in concert with the generation of third-harmonic light within the device. This effect blue-shifted the cavity resonance frequency by $20.2\,(2)\,\text{GHz}$, exceeding the cavity linewidth.
The shift corresponded to a fractional change in refractive index of $-1.05\,(1)\times10^{-4}$, and its relaxation occurred over several tens of hours.
Although photorefraction is a second-order nonlinear effect and has previously not been observed in diamond owing to its vanishing $\chi^{(2)}$, the observed behaviour is consistent with the generation of non-zero $\chi^{(2)}$ by electric fields from charged crystal defects. 
In addition to allowing tuning of diamond cavities for resonant nonlinear and quantum photonics applications, this observation could enable the realisation of diamond frequency converters and electro-optical modulators that rely on second-order nonlinearity.
}

\begin{document}

\preprint{AIP/123-QED}

\title{\doctitle}
\author{\authorOne}
\affiliation{\addressOne}%
\author{\authorTwo}
\affiliation{\addressOne}%
\affiliation{\addressTwo}%
\affiliation{\addressThree}%
\author{\authorThree}
\affiliation{\addressOne}%
\author{\authorNine}
\affiliation{\addressOne}%
\author{\authorFour}
\affiliation{\addressOne}%
\author{\authorFive}
\affiliation{\addressOne}%
\author{\authorSix}
\affiliation{\addressOne}%
\affiliation{\addressThree}%
\affiliation{\addressFour}%
\author{\authorSeven}
\email{\emailContactOne}
\affiliation{\addressOne}%
\author{\authorEight}
\email{\emailContact}
\affiliation{\addressOne}%

\begin{abstract}
\abstractText
\end{abstract}

\maketitle
\section{Introduction}
Efficient enhancement of light-matter interaction is a cornerstone for many applications in nonlinear and quantum optics. Prominent examples include frequency conversion processes that rely on second- ($\chi^{(2)}$) and third- ($\chi^{(3)}$) order nonlinear interactions, which have enabled new lasers at exotic wavelengths~\cite{Mildren2009}, and single photon frequency conversion~\cite{Dreau2018}, an important tool for creation of quantum networks~\cite{Knaut2024,Stolk2024}. 
The efficiency of these processes hinges on the inherently weak nature of nonlinear light-matter interactions.
Fortunately, nonlinear processes can be greatly enhanced by tightly confining light in optical resonators, which has been successfully demonstrated for various materials, including but not limited to diamond~\cite{Hausmann2014,Latawiec2018,Flagan2025}, gallium phosphide~\cite{Lake2016,Logan2018,Wilson2020}, silicon nitride~\cite{Lu2021,Clementi2025} and lithium niobate~\cite{Chen2019Optica,Sun2017OpticsExpress,Boes2023}.

The field of diamond photonics~\cite{Aharonovich2011NaturePhot,Schroder2016,Janitz2020,Shandilya2022} has grown rapidly in recent years, owing to diamond's excellent optical properties and innovations in fabrication techniques~\cite{Maletinsky2012,Burek2012,Khanaliloo2015NanoLett,Hedrich2020,Ding2024,Jing2024Nature,Riedel2026}.
Diamond possesses a moderate refractive index ($n_{0}\simeq2.4$) and a comparatively large ${\chi^{(3)}\sim10^{-22}\,\text{m}^2\text{V}^{-2}}$\,\cite{Hausmann2013DiamondResonator,Hausmann2014,Ososanya2025}, where the exact value depends on the concentration of crystal defects\,\cite{Abulikemu2023,Talik2025}. 
Furthermore, its large bandgap ($5.5\,\text{eV}$) and correspondingly wide transparency window mitigate two-photon absorption, which, when combined with the high thermal conductivity, allow diamond photonic devices to support large optical intensities without sustaining optical damage or exhibiting nonlinear optical absorption. 
These physical properties have enabled the demonstration of third-harmonic generation (THG)~\cite{Abulikemu2023,Zheng2024PRApplied}, four-wave mixing~\cite{Rand1988,Hausmann2014}, Kerr effects~\cite{Almeida2017,Shalaby2017,Motojima2019,Behjat2023,Talik2025}, and Raman scattering~\cite{Lee2011,Latawiec2015,Latawiec2018,Anderson2018,Riedel2020,Flagan2022Dres} from bulk diamond and photonic devices fabricated from single-crystal diamond. 
Such phenomena make diamond a powerful material in the field of photonics~\cite{Aharonovich2011NaturePhot,Schroder2016,Janitz2020,Shandilya2022}.

In addition to being a promising material platform for photonic devices, the diamond lattice hosts various defect centres~\cite{Zaitsev2010,Hepp2014,Iwasaki2017,Rose2018,Trusheim2019,Mukherjee2023,Corte2023,Wang2024PRL,Morris2025}. 
Considerable attention has been paid to nitrogen-related defects~\cite{Ashfold2020}, such as substitutional nitrogen (N$_\text{s}$)~\cite{Cox1994,Ferrari2018} and nitrogen-vacancy (NV) centres~\cite{Doherty2013}. 
Although their internal spin- and optical properties have been studied in detail~\cite{Maze2011,Aslam2013,Chu2015,Baier2020,Degen2021,Razinkovas2021Photoionization,Blakley2024}, the effect of these defects on the nonlinear optical properties of the host diamond crystal is not well-understood~\cite{Almeida2017,Motojima2019,Talik2025}. 
For example, the local electric field in a crystal depends on the spatial configuration and occupation of charged defects~\cite{Yurgens2021}. 
The optical excitation and photoionisation of these defects alter the local charge environment and can manifest in the formation of space-charge distributions~\cite{Goldblatt2026}. 
A well-established consequence of space-charge redistribution is the photorefractive effect, which has been routinely observed in lithium niobate~\cite{Ashkin1966APL,Glass1974,Savchenkov2006,Leidinger2016,Liang2017Optica,Jiang2017OptLett,Li2019Optica,Kong2020AdvMat,Zhu2021,Liu2021PRL,Xu2021OpticsExpress,Hou2024LaserReview,Ren2025}. 
Within this effect, the space-charge redistribution creates an electric field, $E_{\text{sp}}$~\cite{Gunter1988,Boyd2008}, which modulates the refractive index via the electro-optic effect~\cite{Xu2021OpticsExpress}. 
Furthermore, electric fields from charged defects have led to the demonstration of second-harmonic generation in silicon waveguides~\cite{Castellan2019} and more recently in diamond microdisks~\cite{Flagan2025}---an otherwise forbidden process in these centrosymmetric materials.

\begin{figure}
\centering
    \includegraphics[width = \columnwidth]{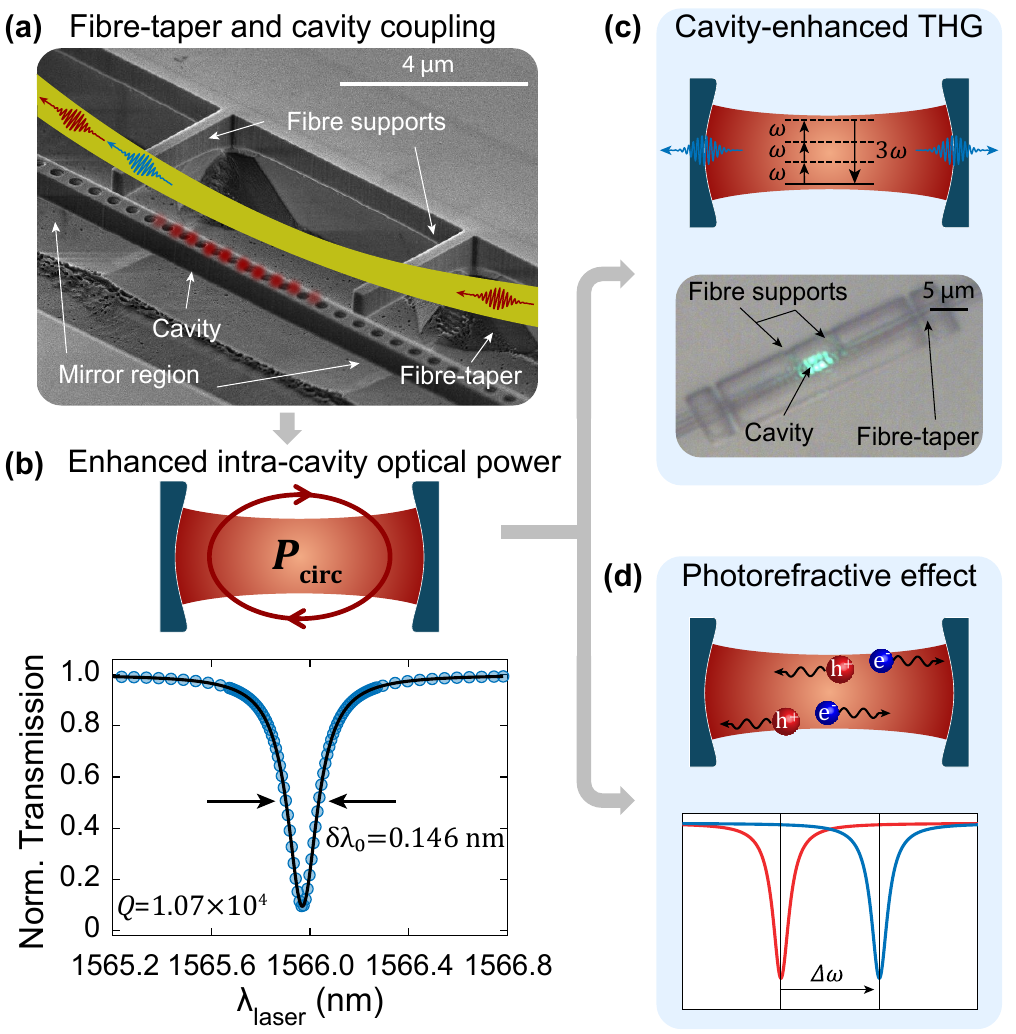}
    \caption{
    \textbf{(a)} Scanning electron micrograph of the diamond photonic crystal cavity, superimposed by a schematic of the fibre-taper and the cavity mode field profile. The fibre-taper rests on the fibre supports patterned adjacent to the device to fix it at a constant distance from the device. 
    \textbf{(b)} Bottom panel: The blue data points show the transmission through the fibre-taper as the laser scans across the cavity resonance for low input power. A cavity mode is observed at $\lambda_0=1565.98\,(1)\,\text{nm}$ with a resonance contrast of $\sim90\,\%$. The black line shows a Lorentzian fit used to extract the loaded and intrinsic $Q$ factors of $1.07\,(1)\times 10^4$ and $3.47\,(9)\times10^{4}$, respectively. Top panel: On resonance, the optical power inside the cavity increases due to resonant recirculation.
    \textbf{(c)} Free-space emission of green light, attributed to THG, captured on a CCD camera when the pump laser is near resonance and enhanced by the cavity mode. 
    \textbf{(d)}
    Photo-induced excitation and optical redistribution of charges by the large intracavity intensity lead to the generation of a space-charge electric field, which modulates the refractive index via an induced electro-optic effect leading to a net blue shift of the cavity resonance by $\Delta\omega$. For further discussion of the induced electro-optic effect, see Section 6 of the Supplementary Material.
    }
    \label{Fig:Fig_1}
\end{figure}

In this work, we investigate nonlinear optical phenomena in the fibre-taper-coupled diamond photonic crystal nanocavity shown in the scanning electron micrograph in Fig.\,\ref{Fig:Fig_1}\,(a).
The observation of these inherently weak nonlinear effects is facilitated by the tight confinement of the light field by the nanocavity, enabled by the large ratio of the device's optical quality ($Q$) factor to its optical mode volume ($V$).
Here, the large $Q$-factor enhances the intracavity circulating power, while the small mode volume enhances the optical energy density per intracavity photon~\cite{Carmon2007} (Fig.\,\ref{Fig:Fig_1}\,(b)). 
In particular, using continuous wave lasers at moderate power levels ($\sim$mW), we demonstrate cavity-enhanced third-harmonic generation that converts three telecom wavelength infrared (IR) photons to green emission, which is visible when imaging the diamond device on a CCD camera (Fig.\,\ref{Fig:Fig_1}\,(c)).
Furthermore, we observe that prolonged input of IR light into the cavity blue-shifts its resonance frequency. 
We attribute this blue-shifting to photorefraction~\cite{Li2019Optica}, a novel effect in diamond cavities that is related to the presence of defect centres, and we demonstrate that its observation hinges upon the presence of green light. Relaxation or subsequent red-shifting of the cavity can be initiated by substantially reducing the input power, resulting in a slow relaxation process over several tens of hours.

The potential impact of photorefractive tuning of diamond cavity resonances is substantial. 
For example, it offers a robust non-volatile method for aligning optical modes with the electronic transition frequency of embedded single-photon emitters. 
This alignment method could enable stable \textit{in situ} cavity resonance tunability of individual devices on the same chip without degrading the cavity quality factor.
Such control would be valuable for quantum applications involving Purcell enhanced optical coupling to colour centres\,\cite{Bhaskar2020,Knaut2024}, improving the sensitivity of absorption-based magnetometers\,\cite{Jensen2014,Schall2025}
or for realising doubly-resonant cavities for nonlinear optics applications, where tunability could enhance the efficacy of technologies such as Brillouin and Raman lasers\,\cite{Flagan2022Dres}.

In the next section (\Cref{sec:PCC}), we describe the photonic crystal cavity studied here, including details of its fabrication, optical properties, and the experimental methods used to characterize it. 
\Cref{sec:THG} details the third-harmonic light generated in the photonic crystal cavity, followed by \Cref{sec:tuning}, which discusses the photorefractive blue-shifting of the cavity resonance. 
Finally, \Cref{sec:applications} reviews some applications of the observed effects. 
We include a Supplementary Material as an addendum to this manuscript, which provides additional details on the experiments and models used to analyse the third-harmonic generation and photorefractive blue-shifting.

\section{Photonic Crystal Cavity}
\label{sec:PCC}
The photonic crystal cavity (PCC) studied in the work is shown in the scanning electron micrograph in Fig.\,\ref{Fig:Fig_1}\,(a)  and was fabricated from single-crystal diamond (Element Six, `optical grade') using the previously reported quasi-isotropic etching process~\cite{Khanaliloo2015NanoLett, Mitchell2019}, with an improved Si-rich SiN$_{\text{x}}$ hard mask for better sidewall protection. 
The PCC consists of a suspended diamond nanobeam patterned with holes whose nominal periodic spacing and diameter vary in the cavity region, so that it supports a localised optical mode with a wavelength near 1550\,nm. 
The cavity design details and mode properties are presented in the Supplementary Material.  
To load photons into the cavity, the PCC is evanescently coupled to the guided mode of a fibre-taper waveguide.
A constant fibre-to-device distance is maintained by positioning the fibre-taper on the fibre supports patterned near the cavity, as shown in Fig.\,\ref{Fig:Fig_1}\,(a)~\cite{Masuda2024}. 
To probe the PCC optical modes, we inject the output of a widely tunable CW laser into the fibre-taper, and measure its transmission as a function of wavelength. 
We find a cavity mode centred at $\lambda_0 = 1565.98\,(1)\,\text{nm}$ with a full-width at half-maximum linewidth of $\delta\lambda_0 = 0.146\,(2)\,\text{nm}$ ($\delta\omega_0/2\pi=17.8\,(2)\,\textrm{GHz}$).

The coupling between the fibre-taper waveguide and the PCC can be described using coupled-mode theory~\cite{Aspelmeyer2014,Sun2017OpticsExpress}, from which an expression for the normalised fibre-taper transmission can be obtained: 
\begin{equation}
        T = \left|e^{i\phi_1} - \frac{\kappa_{\text{1,ex}}}{-i\Delta+ \frac{\kappa_1}{2}}\right|^2\,.
    \label{eq:General_Transmission}
\end{equation}
Here, $\Delta$ is the detuning between the drive laser at frequency $\omega$ and the cavity resonance at frequency ${\omega_0 = 2\pi c/\lambda_0}$, and $\kappa_\text{1,ex}$ is the coupling decay rate of the cavity. 
Fano interference effects between the transmitted input and coupled cavity fields are responsible for certain asymmetries in the cavity line shape and are modelled using phase $\phi_1$\,
\cite{Fan2003}.
The total decay rate is given by $\kappa_1 = \kappa_\text{1,i+p} + 2\kappa_\text{1,ex}$, where $\kappa_{1,\text{i+p}}$ encompasses the intrinsic cavity loss and parasitic loss introduced by the fibre-taper~\cite{Barclay2005}, respectively. 
The coupling decay rate of the cavity contributes twice because the standing wave mode couples equally well to forward- and backward propagating modes in the fibre-taper.
For a cold cavity (with low laser input power), we have ${\Delta = \Delta_0 = \omega - \omega_0}$. However, as we will show below, the functional form of $\Delta$ changes in the presence of thermo-optic and photorefractive effects.  
By fitting the transmission spectrum using Eq.\,\ref{eq:General_Transmission}, we extract $\kappa_1$, $\kappa_{\text{1,i+p}}$ and $\kappa_{\text{1,ex}}$, which respectively yield the total, intrinsic and parasitic, and coupling quality factors, $Q=\omega_0/\kappa_1$, $Q_{\text{i+p}}=\omega_0/\kappa_{1\text{,i+p}}$ and $Q_{\text{ex}}=\omega_0/\kappa_{1\text{,ex}}$. 
We list these parameters in Tab.\,\ref{table:cavity_parameters}. 
From finite element simulations of the PCC optical mode, we calculate its mode volume, $V_{\text{eff}}$, and deduce ${Q}/{V_{\text{eff}}}\sim10^{5}\times\left(\,{\lambda_0}/{n_0}\right)^{-3}$ (see the Supplementary Material). 

A key property of diamond nanophotonic devices is their ability to support intense fields at telecom wavelengths without the onset of multiphoton absorption. 
In our device, at the maximum experimental power input into the fibre-taper, $P_{\text{in}}=75\,\text{mW}$,
the intracavity photon number exceeds three million -- a result of the large $Q$ factor. 
The resulting combination of large intracavity photon number and small mode volume manifests in a high intracavity peak field intensity $I_\text{peak} = 56\,\text{GW}/\text{cm}^2$ (see the Supplemental Material), illustrating how the PCC's large ${Q}/ {V_{\text{eff}}}$ and intracavity photon number makes it an ideal platform for exploring nonlinear interactions~\cite{Carmon2007}. 
This is, to the best of our knowledge, the largest field intensity reported in a diamond nanophotonic cavity using a CW laser.
We note that in this experiment we limited the laser power to $75\,\text{mW}$ to avoid thermal damage to the fibre-taper.

\begin{table}[t]  
\centering 
\caption{Summary of cavity parameters for low input power. Uncertainties were estimated from the residuals of the Jacobian matrix obtained in the least-squares fit.}
\begin{tabularx}{1\columnwidth}{l @{\extracolsep{\fill}} l l l} 
\hline\hline   
Parameter                       & \multicolumn{1}{c|}{Value}                            & Parameter                 & \multicolumn{1}{c}{Value}\\
\hline \hline 
$\lambda_0$                         & \multicolumn{1}{c|}{{$1565.98\,(1)\,\text{nm}$}}      & $\delta\lambda_0$         & \multicolumn{1}{r}{$0.146\,(2)\,\text{nm}$} \\
$\omega_0/ 2\pi$                    & \multicolumn{1}{r|}{{$ 191.441\,(1)\,\text{THz}$}}    & $\delta\omega_0/ 2\pi$    & \multicolumn{1}{r}{$17.8\,(2)\,\text{GHz}$} \\
$\kappa_\textrm{1,i+p}/2\pi$        & \multicolumn{1}{r|}{$5.5\,(1)\,\text{GHz}$}           & $Q_{\text{i+p}}$          & \multicolumn{1}{r}{$3.47\,(9)\times10^4$ } \\
$\kappa_{\text{1,ex}}/ 2\pi$        & \multicolumn{1}{r|}{$6.15\,(5)\,\text{GHz}$}          & $Q_{\text{ex}}$           & \multicolumn{1}{r}{$3.11\,(3)\times 10^4$} \\
$\kappa_{\text{1}}/ 2\pi$           & \multicolumn{1}{r|}{$ 17.8\,(2)\,\SI{}{GHz}$}         & $Q$                       & \multicolumn{1}{r}{$1.07\,(1)\times 10^4$} \\
$V_{\text{eff}}$ \vspace{0.02cm}    & \multicolumn{1}{r|}{$0.13\times\left(\frac{\lambda_0}{n_{0}}\right)^3$} & $Q/V_{\text{eff}}$  & \multicolumn{1}{r}{$10^5\times\left(\frac{\lambda_0}{n_{\text{0}}}\right)^{-3}$}\\
\hline 
\hline 
\label{table:cavity_parameters}
\end{tabularx} 
\end{table} 

\section{Cavity-Enhanced Third-Harmonic Generation}
\label{sec:THG}
\begin{figure}[bt!]
\centering
    \includegraphics[width=\columnwidth]{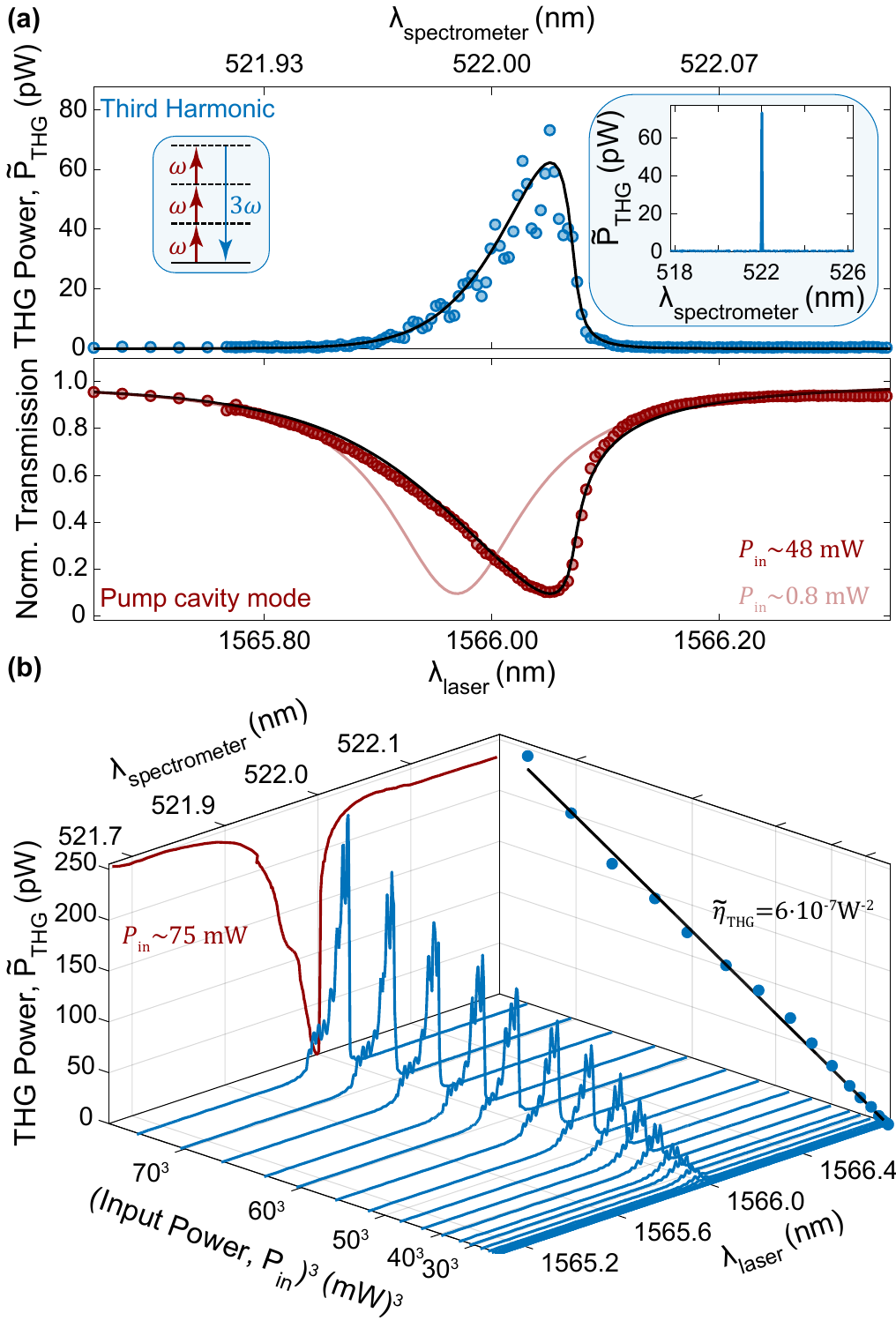}
\caption{Cavity-enhanced third-harmonic generation from a diamond photonic crystal cavity.
\textbf{(a)} Fibre-taper transmission (bottom panel) and the corresponding THG output power (top panel) as the laser is step-wise tuned across the cavity resonance for $P_{\textrm{in}}\sim48\,\textrm{mW}$. For this large input power, the cavity line shape deviates from a Lorentzian due to the thermo-optic effect. For comparison, the faint red line shows the cold-cavity line shape for $P_{\textrm{in}}\sim0.8\,\textrm{mW}$. The solid black lines are fits to the data. For details, see the main text. The inset in the top panel shows the THG spectrum for $\lambda_{\text{laser}}=1566.05\,\text{nm}$.
\textbf{(b)} The detuning dependence of the THG emission spectrum with increasing input laser power. Note that the $y$-axis shows input power cubed, i.e. $P_{\text{in}}^3$. The red projection shows that for large power ($P_{\text{in}}\sim 75\,\textrm{mW}$), the cavity transmission spectrum becomes distorted due to optomechanical induced self-oscillation, as manifested by the shoulder on the blue side of the cavity mode. The observed optomechanical self-oscillation is also evident in the THG spectra. The blue projection shows the peak THG count for each input power cubed. The black line is a linear fit, yielding an end-to-end conversion efficiency $\tilde{\eta}_\textrm{THG}=6\cdot10^{-7}\,\textrm{W}^{-2}$.
}
\label{Fig:THG_PD}
\end{figure}

We now demonstrate how cavity-enhanced third-harmonic generation can be observed from this system. We start by step-wise tuning the wavelength of the pump laser across the cavity mode while simultaneously recording the resulting pump laser transmission and THG emission through the fibre-taper using a photodiode and spectrometer, respectively. 
An example of this measurement is shown in Fig.\,\ref{Fig:THG_PD}(a), where the bottom panel shows the dependence of the fibre-taper transmission on wavelength for $P_{\text{in}}\sim45\,\textrm{mW}$. 
Compared to Fig.\,\ref{Fig:Fig_1}\,(b), the elevated IR power leads to an asymmetric cavity line shape due to the thermo-optic effect, which occurs when an optical cavity undergoes laser-induced heating, leading to a modification of the refractive index~\cite{Schmidt2008}. The time scale for this effect is $\sim\upmu\text{s}$\,\cite{Lake2018}, which is much faster than the laser sweep rate (${\sim\text{s}}$)~\cite{Carmon2004}, and when combined with thermal expansion, which effectively enlarges the cavity length, the thermo-optic effect red-shifts the cavity resonance, resulting in an asymmetric cavity line shape~\cite{Barclay2005,Gupta2007,Shankar2011}.  
This effect is well-described by replacing $\Delta$ in Eq.\,\ref{eq:General_Transmission} with ${\Delta_{\text{TO}}=\Delta_0-c_\text{T}N}$, where $c_\text{T}$ is the thermo-optic coefficient~\cite{Hu2021PRL, Sun2017OpticsExpress}. 
The intracavity photon number, $N = N(\Delta_0,P_\text{in})$, depends on both the input laser power $P_{\text{in}}$ and the detuning $\Delta_0$. 
The term $c_{\text{T}}N$ captures the thermo-optic effect and describes the nonlinear red shift of the cavity resonance and the deviation from a Lorentzian line shape (see the Supplementary Material). 

For $P_{\text{in}}\sim70\,\text{mW}$, we observe that the thermo-optic effect red shifts the cavity resonance by ${\Delta\omega_{\text{TO}}/2\pi=-9.1 (2)\,\text{GHz}}$ (Fig.\,\ref{fig:PR_effect}\,(a)), corresponding to a change in cavity temperature of $\Delta T\sim12\,\text{K}$ (see the Supplemental Material).
We note that although the cavity line shape exhibits photo-thermal behaviour, the increase in temperature is not sufficiently large to observe thermal bistability: we do not observe a significant hysteresis in the cavity line shape between forward and backward wavelength sweeps. Furthermore, as we will elucidate below, we note that the small increase in temperature is unlikely to cause permanent thermal damage to the diamond nanocavity. 

Next, we analyse the dependence of the fibre-taper-collected THG signal on the pump wavelength. 
As shown in the top panel of Fig.\,\ref{Fig:THG_PD}\,(a), the THG signal is largest when the optical power dropped into the cavity is maximum, as expected. 
We also observe that the thermo-optic effect's distortion of the cavity transmission line shape is imprinted on the pump detuning dependence of the THG signal. 
It is important to note that the PCC studied here was not designed to support any confined cavity modes in the visible wavelength range of the THG signal. 
Instead, the device behaves like a waveguide for the third-harmonic-generated light at frequency $3\omega_0$. 
However, to facilitate the modelling of our system, we treat the waveguide as a weak cavity with total and coupling decay rates $\kappa_3$ and  $\kappa_{3,\text{ex}}$, respectively.
Note that in nanophotonic devices like the cavity presented in this work, the importance of phase-matching is relaxed by the localised nature of the spatial overlap between the fundamental and harmonic fields\,\cite{Rodriguez2007,Rivoire2010}.

The THG signal shown in Fig.\,\ref{Fig:THG_PD}\,(a) can be described using the thermo-optic model for the cavity transmission to determine the power dropped into the cavity, from which the following dependence of the THG output power, $\tilde{P}_{\text{THG}}$ on the input power $P_{\text{in}}$ and detuning $\Delta_0$ can be derived\,\cite{Galli2010, Lake2016, Hu2021PRL, Sun2017OpticsExpress, Mclaughlin2022} (see the Supplementary Material):
\begin{equation}
    \tilde{P}_{\text{THG}} =  \frac{108}{\hbar^2}\times\mathcal{L}\times\left[\frac{\kappa_{1,\text{ex}}}{\Delta_\textrm{TO}^2 + \left(\frac{\kappa_1}{2}\right)^2}\right]^3{\left(\sqrt{{\eta}_\text{fibre}}P_{\text{in}}\right)^3}\,.
    \label{eq:THG}
\end {equation}
Here, $\mathcal{L} = |\beta_\textrm{THG}|^2\times\left(\kappa_{3,\textrm{ex}}/{\kappa_3^2}\right)$, where $\beta_{\text{THG}}$
is a constant that describes the inter-modal overlap between the pump cavity mode and the THG waveguide mode, and $\eta_{\text{fibre}}$ is the transmission efficiency of the tapered fibre at 1565\,nm, which we assume is equally divided between the fibre-taper regions before and after the cavity.
We are unable to perform coherent mode spectroscopy to independently measure $\kappa_{3}$ and $\kappa_{3,\text{ex}}$ as we do not have access to a tunable laser at the THG wavelength.
By fitting Eqs.\,\ref{eq:General_Transmission} and \ref{eq:THG} to transmission and THG spectrums, respectively, (black lines in Fig.\,\ref{Fig:THG_PD}\,(a)) we extract 
$c_{\text{T}}/ 2\pi=-4.53\,(7)\,\text{kHz}$ and $\mathcal{L} = 2.9\,(3) \times 10^{-44}\,\text{s}$.  We find good agreement between our model and the THG signal, confirming that the emission depends cubically on intracavity power, as expected for a third-order process. To further test this dependence, we study the variation of the THG signal with input power. To this end, we repeat the wavelength detuning sweep, with $P_{\text{in}}$ ranging from $0.8\,\text{mW}$ to $75\,\text{mW}$, as measured immediately before the fibre-taper.
The resulting spectra, shown in Fig.\,\ref{Fig:THG_PD}\,(b), exhibit the photo-thermally modified line shapes described above, as well as a signature of mechanical self-oscillations at high input power~\cite{Shandilya2021}, manifested by the emergence of a shoulder on the blue side of the cavity resonance~\cite{Krause2015}.
Also plotted in \Cref{Fig:THG_PD}\,(b) is the maximum THG output power achieved when $\Delta_{\text{TO}} = 0$. 
From Eq.\ \ref{eq:THG}, we see that $P_{\textrm{THG}}^\text{max}\propto P_{\textrm{in}}^3$, which is in good agreement with the data.
For the largest input power, we estimate the circulating third-harmonic generated power inside the cavity to be in the range of $3\dots480\,\text{nW}$ (see Supplemental Material section\,5B).

From these measurements of $P_{\textrm{THG}}^\text{max}\left(P_\textrm{in}\right)$,  we can estimate the end-to-end conversion efficiency, $\tilde{\eta}_{\text{THG}}$. After calibrating the losses in the fibre link to extract the THG power at the fibre-taper output, $\tilde{P}_{\text{THG}}$ (see the Supplementary Material),  we find ${\tilde{\eta}_\textrm{THG}=\tilde{P}_{\text{THG}}/{P_{\text{in}}^3}=6\cdot10^{-7}\,\text{W}^{-2}}$ from the linear fit of $\tilde{P}_{\textrm{THG}}$ as a function of $P_{\textrm{in}}^3$. In this estimation, we have not corrected for poor coupling between the guided mode in the fibre-taper and the third-harmonic generation. 
The conversion efficiency could be improved by using a dedicated fibre-taper with better phase matching to the frequency-converted light~\cite{Logan2018,Wang2022}.
Finally, we note that both the fibre-taper and the first 90:10 beam splitter are designed for operation at telecom wavelengths. We therefore attribute the noisy oscillations observed in the THG spectra to etaloning occurring as an artefact of using the inadequate beam splitter for the THG wavelength---a dedicated fibre-taper for the THG wavelength would mitigate these oscillations. Similar oscillations have also been observed in experiments on harmonic generation in other fibre-taper-based systems\,\cite{Lake2016}.

\section{Photorefractive Blue-shifting}
\label{sec:tuning}
\begin{figure*}[t!]
\centering
    \includegraphics[width = \textwidth]{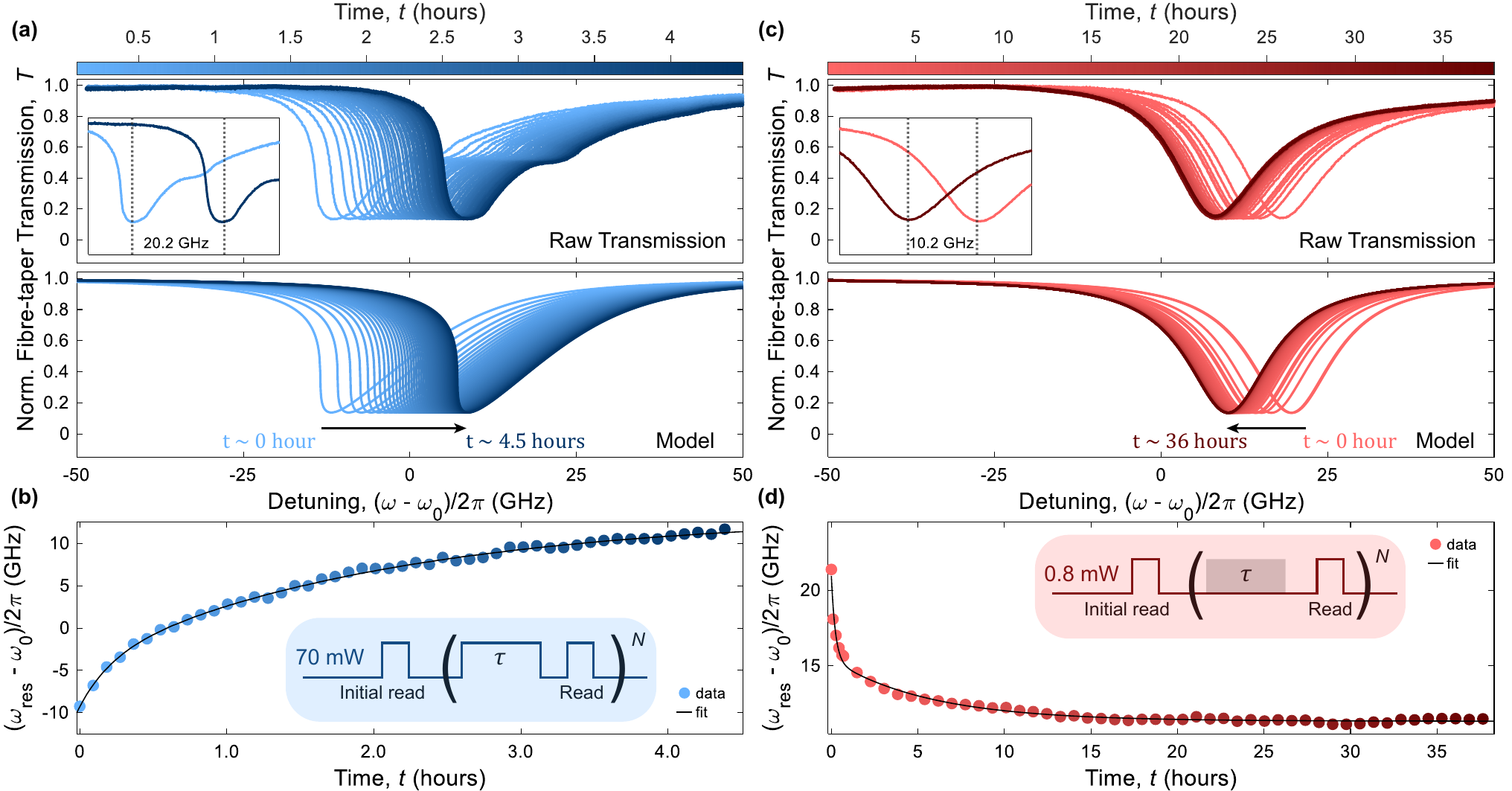}
    \caption{Simultaneous thermo-optic and photorefractive effects. \textbf{(a)} The top panel shows the normalised raw fibre-taper transmission scans for $P_{\text{in}}\sim70\,\textrm{mW}$. The laser is parked on resonance for five minutes between each laser scan, and, during this time, the cavity resonance blue-shifts due to photorefraction. The inset shows that the resonance contrast and cavity line shape remain constant, thus excluding blue-shifting due to movement of the fibre-taper.
    The colour gradient of each scan shows the elapsed time of the repeated high-power scans. 
    After a total duration of $\sim4.5$\,hours, the cavity mode blue-shifts by $\Delta\omega/ 2\pi=20.2\,(2)\,\text{GHz}$ (inset), exceeding the cold-cavity linewidth.
    The middle panel shows the theoretical transmission spectrum calculated from our model. The initial red shift is a consequence of the thermo-optic effect.
    \textbf{(b)} The blue data points show the blue shift of the cavity resonance due to photorefraction with elapsed time, with the black line being the fit to the model.
    \textbf{(c)} Red-shifting of the cavity resonance due to photorefraction relaxation. The top panel shows the raw fibre-taper transmission as the laser is repeatedly scanned across the cavity resonance for low input power. After a duration of $\sim36$\,hours, the cavity red-shifts by $10.2\,(2)\,\text{GHz}$ towards the cold-cavity resonance. Similarly to (a), the middle panel shows the theoretical transmission spectrum calculated from our model.
    \textbf{(d)} The red data points show the red-shifting of the cavity resonance due to photorefractive relaxation with elapsed time. The black line is a fit to the model. For details, see the main text.
    The insets in (b) and (d) describe the experimental procedure used for each measurement, where $\tau=5$\,minutes is the waiting time between each transmission sweep.
    We repeat each procedure $N$ times.
    }
    \label{fig:PR_effect}
\end{figure*}
While performing the THG measurements, we observed an unexpected gradual blue shift of the cavity resonance. 
To investigate this frequency shifting in more detail, we measured the cavity mode wavelength over long periods while driving the cavity with a high input power ($P^{\textrm{high}}_{\text{in}}\sim70\,\text{mW}$). 
These measurements aimed to gain insight into the temporal behaviour of the blue-shifting effect.

The measurement sequence started with a fast laser transmission scan at low power to record the cavity resonance frequency in the absence of thermo-optic effects. 
We then performed a fast scan at high input power to determine the thermo-optically shifted cavity resonance frequency, after which we fixed the laser wavelength onto resonance for $\tau=5$\,minutes. The prolonged high-power laser exposure blue-shifted the cavity resonance frequency, which we immediately measured using another fast laser transmission scan. 
The top panel in Fig.\,\ref{fig:PR_effect}\,(a) shows a series of 50 high input power transmission scans. 
Over the duration of the measurements ($\sim5\,\text{hours}$), the resonance contrast was unchanged, as observed in similar experiments using lithium niobate~\cite{Li2019Optica}. 
This confirms that the observed blue-shifting of the cavity resonance is related to a purely dispersive change in cavity properties, and is not related to movement of the fibre-taper relative the PCC, which would also change the cavity linewidth and contrast. 
In Fig.\,\ref{fig:PR_effect}\,(b), we plot the cavity resonance frequency shift with respect to the cold-cavity resonance frequency extracted from Fig.\,\ref{Fig:Fig_1}\,(b) as a function of time ($t$). We observe a blue shift of the cavity resonance that slows with time.
The initial (${t = 0}$) negative resonance shift of ${\Delta\omega_{\text{TO}}/2\pi=-9.1\,(2)\,\text{GHz}}$ is a result of the aforementioned thermo-optic effect, which induces a red shift due to heating induced by the high laser power. After 4 hours, we observe a total cavity blue shift of ${\Delta\omega/ 2\pi = 20.2\,(2)\,\text{GHz}}$. We note that this shift exceeds the cold-cavity linewidth,
${{\delta\omega_0}/{2\pi}={c\delta\lambda_0}/{\lambda_0^2}=17.8\,(2)\,\textrm{GHz}}$, extracted from Fig.\,\ref{Fig:Fig_1}\,(b).
Also note that we initially park the laser on resonance; however, during the driving period, $\tau$, the resonance blue shifts away from the pump laser, meaning that the intracavity power varies during the driving period.
This variability in intracavity power is not significant once the tuning slows; near the start, however, a sufficiently short driving period is needed so that the pump laser stays close to resonance.
From the measured frequency blue shift of $\Delta\omega/ 2\pi =20.2\,(2)\,\text{GHz}$ and by using the relationship
${\Delta\omega}/{\omega_0}\simeq-{\Delta n}/{n_0}$~\cite{Savchenkov2006}, we calculate ${\Delta n}/{n_0}=-1.05\,(1)\times10^{-4}$. 
To the best of our knowledge, we have measured the largest refractive index modulation in a diamond cavity, which is comparable to what has been previously demonstrated in more mature electro-optic materials, such as lithium niobate~\cite{Peithmann1999,Savchenkov2006}, lithium tantalate~\cite{Holtmann2004,Liu2004JAP}, and barium titanate~\cite{Ortmann2019,Karvounis2020}.
Finally, to verify the reproducibility of the resonance tuning, we perform a control experiment composed of an alternating measurement sequence of high and low IR power laser scans. Blue- and red-shifting is consistently observed for high and low power exposure, respectively. This verification of our observation, explained in detail in the Supplemental Material, demonstrates that careful toggling of laser power and exposure enables deterministic \textit{in situ} control of the cavity resonance frequency.

\setlength{\extrarowheight}{2pt}
\begin{table}[t]  
\centering 
\caption{Summary of photorefractive shift parameters.}
\begin{tabularx}{1\columnwidth}{l @{\extracolsep{\fill}} c c c} 
\hline\hline   
Parameter & \multicolumn{1}{c|}{Value} & Parameter & Value \\ \hline \hline 
$\Delta\omega/ 2\pi$                    & \multicolumn{1}{r|}{$20.2\,(2)\,\si{GHz}$}    & $\frac{\Delta n}{n_0}$    & \multicolumn{1}{r}{$-1.05\,(1)\times10^{-4}$}  \\
$C_{\text{f}}^{\text{PR}}/ 2\pi$        & \multicolumn{1}{r|}{$8\,(1)\,\si{GHz}$}       & $\Gamma_{\text{f}}$       & \multicolumn{1}{r}{$5\,(1) \,\text{hours}^{-1}$}  \\
$C_{\text{s}}^{\text{PR}}/ 2\pi$        & \multicolumn{1}{r|}{$18.2\,(6)\,\si{GHz}$}    & $\Gamma_{\text{s}}$       & \multicolumn{1}{r}{$0.48\,(7)\,\text{hours}^{-1}$}  \\
$\tilde{C}_{\text{f}}^{\text{R}}/ 2\pi$ & \multicolumn{1}{r|}{$5.5\,(2)\,\si{GHz}$}     & $\gamma_{\text{f}}$       & \multicolumn{1}{r}{$4.3\,(4)\,\text{hours}^{-1}$}  \\
$\tilde{C}_{\text{s}}^{\text{R}}/ 2\pi$ & \multicolumn{1}{r|}{$4.1\,(1)\,\si{GHz}$}     & $\gamma_{\text{s}}$       & \multicolumn{1}{r}{$0.177\, (7) \,\text{hours}^{-1}$}  \\
\hline 
\hline 
\label{table:fitting_parameters}
\end{tabularx} 
\end{table} 

\subsection{Photorefraction mechanism}
We suggest thatthe blue-shifting of the cavity resonance arises from the photorefractive effect\,\cite{Gunter1988}, on account of the close resemblance to results obtained in other systems such as lithium niobate devices\,\cite{Sun2017OpticsExpress,Jiang2017OptLett,Ren2025}.
We support this hypothesis on several grounds.
First,  the relaxation indicates a temporary modification to the cavity, and we can therefore exclude permanent photo-thermal damage by the intense intracavity field.
Second, while high-power laser irradiation is routinely used to create defects in diamond\,\cite{Chen2017,Yurgens2021,Guo2024APL_Phot,Addhya2024,Zhou2024NatPhot}, the intensity threshold required to create single vacancies (GR1 centres) is three orders of magnitude larger than the peak field intensity in our work\,\cite{Zhou2024NatPhot}.  
Alternatively, the formation of graphite on the surface due to prolonged heating by the cavity field is excluded on the account that the change in device temperature is $\sim12\,\text{K}$, which is far below what is required for graphitisation\,\cite{Zhang2025FuncDiamond}.
Furthermore, surface graphite would lead to red-shifting and a degradation in the $Q$-factor due to increased absorption\,\cite{Kavatamane2026}, neither of which is observed. All of these observations are consistent with photorefraction being the dominant explanation for the blue-shifting observed in this work.

In this model, with the laser on resonance, the strong intracavity optical field facilitated by the large ${Q}/ {V_{\text{eff}}}$ ratio of the PCC leads to photoionisation and diffusion of charges associated with defects in the material, forming a space-charge-induced electric field, $E_{\text{sp}}$~\cite{Gunter1988,Hou2024OpticsLett}.
This electric field modulates the refractive index via the electro-optic effect, causing a net blue shift of the cavity resonance~\cite{Liang2017Optica}.
Owing to its highly localised nature, the cavity mode is only sensitive to changes in the refractive index in the centre region of the device.
The observation of the electro-optic effect---a second-order nonlinear effect---is unexpected in diamond on the account of the centrosymmetric crystal structure and vanishing $\chi^{(2)}=0$. Nevertheless, charged crystal defects generate static electric fields, $E_{\text{DC}}$, which can couple to the bulk $\chi^{(3)}$ to induce an effective $\chi^{(2)}_{\text{eff}}=3\chi^{(3)}E_{\text{DC}}$\,\cite{Castellan2019,Flagan2025}.
This non-zero $\chi^{(2)}_{\text{eff}}$ enables the observation of the electro-optic effect. However, one would also expect surfaces effects\,\cite{Trojanek2010,Levy2011,Zhang2019,Wang2025Optica}, residual strain\,\cite{Anastassakis1971} and the presence of defects\,\cite{Li2022CellReports,Abulikemu2021,Abulikemu2022} to contribute towards $\chi^{(2)}\neq0$.

\begin{figure}
\centering
    \includegraphics[width = \columnwidth]{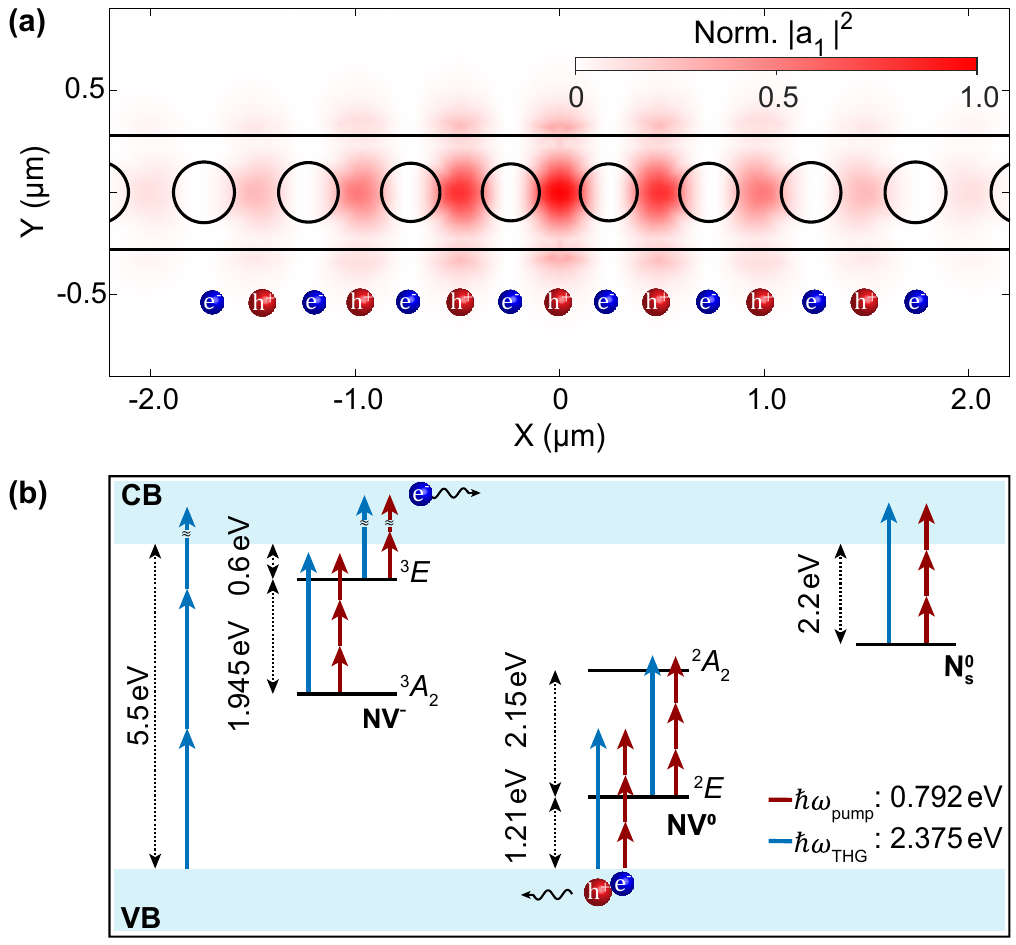}
    \caption{\textbf{(a)} Normalised IR field distribution and the redistribution of charges.
    Electrons liberated by photoionisation processes accumulate in the regions of low intensity. \textbf{(b)} Schematic diagram showing the relative energy levels of the defects within diamond's bandgap. The blue and red arrows indicate the possible photoionisation processes for THG and IR photons, respectively.}
    \label{fig:Fig_4}
\end{figure}

The redistribution of space charges causes modulation of the refractive index due to photorefraction, and a potential mechanism for this effect in the diamond devices studied here is illustrated in Fig.\,\ref{fig:Fig_4}\,(a). 
We speculate that the space-charge redistribution, and consequently the formation of $E_{\text{sp}}$,  arises from optical excitation of crystal defects, such as NV centres, substitutional nitrogen, N$_{\text{s}}$~\cite{Ashfold2020}, and potentially other charge traps on or near surfaces of the PCC~\cite{Bluvstein2019, Stacey2019}, enabled by absorption of the internally generated third-harmonic light~\cite{Leidinger2016}.
We assert the importance of green light below.
For the optical grade diamond samples from which the devices were fabricated, nominal defect concentrations of [N$_{\text{s}}$]$\sim1\,\text{ppm}$ and [NV]$\sim0.01\,\text{ppm}$ are present~\cite{Acosta2009}. 
The relative energy levels of these defects are shown in Fig.\,\ref{fig:Fig_4}\,(b).
Optical excitation of these defects and the resulting liberation of charges can occur directly via either the internally generated third-harmonic field~\cite{Leidinger2016}, multi-photon processes involving the strong IR intracavity field, or processes driven by a combination of the third-harmonic and IR fields\,\cite{Shandilya2024arxiv}.
The two latter mechanisms are restricted to regions of large optical intensity within the centre region of the cavity.
The photoionisation threshold for N$_{\text{s}}$ is $2.2\,\text{eV}$~\cite{Rosa1999,Bourgeois2022}, which can be achieved by the absorption of one THG photon ($\hbar\omega_{522\,\text{nm}}=2.38\,\text{eV}$). 
We anticipate photoionisation of N$_{\text{s}}$ to be the dominant contributor to the charge redistribution due to the large concentration of this defect~\cite{Goldblatt2026}.
In general, photoionisation from the negative to the neutral charge state of the NV centre can occur directly from the $^3A_2$ ground state; however, in our system, this photoionisation process is restricted to occur from the $^3E$ excited state, as direct one-photon ionisation from the ground state is precluded given the available photon energies.
A single THG photon is sufficiently energetic to bring the population from the ground- to the excited state, from where photoionisation can occur either via absorption of a single THG photon or via two-photon IR absorption~\cite{Shandilya2024arxiv}. The latter process is likely given the high IR fields created in the PCC in these studies.
Recombination from the neutral to the negative charge state is possible via a single THG photon (see the Supplementary Material). 
As we alluded to above, we are not able to perform coherent mode spectroscopy to independently measure $\kappa_{3}$ and $\kappa_{3,\text{ex}}$, which precludes reliable conversion of the detected THG signal to intracavity THG field intensity. Therefore, we cannot comment on the ionisation rates that govern the internal charge dynamics and consequently the rate of space-charge formation.
We note that we did not observe any THG-induced NV centre photoluminescence via the fibre-taper, which can be explained by the two-photon IR photoionisation process that becomes dominant for large IR power and traps the population in a dark, non-fluorescent state of NV$^0$~\cite{Shandilya2024arxiv}.
We further note that the effect of the strong IR field on the photophysics of N$_{\text{s}}$ is not known.

\subsection{Photorefraction dynamics}

We now examine the dynamics of the blue-shifting of the cavity resonance in detail. 
We model the fibre-taper transmission by considering the slowly changing photorefractive effect and the quasi-static thermo-optic effect, so that the fibre-taper transmission is obtained by substituting $\Delta$ in Eq.\,\ref{eq:General_Transmission} with $\Delta_\textrm{PR} = \Delta_\textrm{TO} - \Delta_\textrm{PR}(t)$, where the photorefractive cavity resonance frequency shift is given by
$\Delta_{\text{\,PR}}(t) = C_{\text{f}}^{\text{PR}}\left(1-e^{-\Gamma_{\text{f}}\,t}\right) + C_{\text{s}}^{\text{PR}}\left(1-e^{-\Gamma_{\text{s}}\,t}\right)$
(see the Supplementary Material). 
Here, the constants $C_{\text{f,(s)}}$ are the coefficients of photorefraction for a fast (slow) process with characteristic rate $\Gamma_{\text{f,(s)}}$. Note that a multi-exponential response to photorefraction has been observed in lithium niobate, where the fast decay was attributed to the small size of the device and surface effects~\cite{Jiang2017OptLett}. 
We will discuss the use of the bi-exponential model further below.
By fitting $\Delta_\text{PR}(t)$ to the experimentally observed change in resonance frequency with time (Fig.\,\ref{fig:PR_effect}\,(b)), we extract $C_{\text{f}}^{\text{PR}}/ 2\pi = 8\,(1)\,\text{GHz}$, $\Gamma_{\text{f}} = 5\,(1)\,\text{hours}^{-1}$ and ${C_{\text{s}}^{\text{PR}}/ 2\pi = 18.2\,(6)\,\text{GHz}}$, ${\Gamma_{\text{s}} = 0.48\,(7)\,\text{hours}^{-1}}$ for the fast and slow components, respectively.
We find excellent concordance with the shifting line shape data in Fig.\,\ref{fig:PR_effect}\,(a), and we can reproduce the observed saturation in the resonance blue shift at high input power. 
We note that we did not observe any changes to the THG intensity during this measurement.
    
Next, we studied the relaxation of the blue-shifted resonance frequency. 
To this end, we set the laser to low power ($P^{\textrm{low}}_{\text{in}}\sim 0.8\,\text{mW}$) and scanned its wavelength across the cavity resonance once every $\tau=5$\,minutes. 
The recorded cavity line shapes are shown in Fig.\,\ref{fig:PR_effect}\,(c), from which we observe an initial rapid red-shift, which gradually slows down with elapsed time, as shown in Fig.\,\ref{fig:PR_effect}\,(d).
In this measurement, the laser power was sufficiently low to avoid the thermo-optic effect, which is evidently not responsible for the asymmetries present in the measured line shape. 
Instead, the asymmetries are due to Fano interference effects\,\cite{Fan2003}. The same Fano asymmetries present in Fig.\,\ref{fig:PR_effect}\,(c) are present in the high power sweeps (Fig.\,\ref{fig:PR_effect}\,(a)); however, they are less visible due to the thermo-optic effects present in those measurements.
After tracking the cavity resonance over 40 hours, the initial resonance frequency did not fully recover to its initial value.
There are reports of similar observations in bulk glass~\cite{Balakirev1996} and Z-cut lithium niobate microrings~\cite{Ren2025}, where a residual space-charge field is present after the material is exposed to an intense local field. In either case, the charge environment does not fully recover to the initial condition.
Therefore, on the account of the similar behaviour in our system, we suggest that the incomplete relaxation may be related to a similar residual space-charge field within the device. This residual field could arise from several different mechanisms. 
First, the cavity mode is highly localised to the centre region of the device, and only modifications to the refractive index in this region will affect the resonance frequency.
The strong IR field assists in photoionisation\,\cite{Shandilya2024arxiv} and charge diffusion (see \Cref{fig:Fig_4}\,(a)). Charges drifting out of the centre region cannot be recovered, thus leaving behind a residual electric field.
Second, trapping of electrons, either in long-lived surfaces states\,\cite{Ren2025} or in deep-level defects\,\cite{Pu2001,Gorlitz2022} from which they cannot be liberated by the THG light, will prevent recovery of the initial charge environment.
In future experimental studies, recovery of the initial charge environment could potentially be accelerated by surface passivation\,\cite{Salvestrini2011,Guha2017,Holzgrafe2024}, illumination of the device with a high-energy laser, by thermal annealing processes or via a combination of the above. 
In essence, these techniques can facilitate photorefractive relaxation towards the initial resonance frequency.

As with the blue shifting described earlier, the temporal behaviour of the resonance frequency relaxation shown in Fig.\,\ref{fig:PR_effect}\,(d) exhibits two time scales and is well-described by a model in which the laser detuning evolves as $\Delta_\textrm{R} = \Delta_0 - \Delta_\textrm{R}(t)$ in Eq.\,\ref{eq:General_Transmission},
where 
 ${\Delta_{\text{\,R}}(t) = \tilde{C}_{\text{f}}^{\text{R}}\,e^{-\gamma_{\text{\,f}}\,t} + \tilde{C}_{\text{s}}^{\text{R}}\,e^{-\gamma_{\text{\,s}}\,t}}$ (see the Supplementary Material).
By fitting
$\Delta_\text{R}$ to the relaxation data in Fig.\,\ref{fig:PR_effect}\,(d), we extract $\tilde{C}_{\text{f}}^{\text{R}}/ 2\pi = 5.5\,(2)\,\text{GHz}$, $\gamma_{\text{\,f}} = 4.3\,(4)\,\text{hours}^{-1}$ and $\tilde{C}_{\text{s}}^{\text{R}}/ 2\pi = 4.1\,(1)\,\text{GHz}$, ${\gamma_{\text{\,s}} = 0.177\,(7)\,\text{hours}^{-1}}$ for the fast and slow relaxation rates, respectively. 
We use the extracted fit parameters to calculate the expected fibre-taper transmission spectrum (bottom panel, Fig.\,\ref{fig:PR_effect}\,(c)), and find excellent concurrence with the experimental data. 
All experiments were performed using one PCC, though similar behaviour was seen upon repeating the measurements on the same device. 
The absence of complete relaxation indicates a residual charge distribution, precluding exactly repeatable measurements.

The fast and slow time components observed in the photorefraction measurement (Fig.\,\ref{fig:PR_effect}) may be attributed to charge traps associated with the surface and the bulk, respectively, as previously observed in lithium niobate~\cite{Jiang2017OptLett}. 
We expect the photorefractive parameters, $C_{\text{f}}^{\text{PR}}$, $C_{\text{s}}^{\text{PR}}$, $\Gamma_{\text{f}}$ and $\Gamma_{\text{s}}$, to be intracavity power dependent\,\cite{Li2019Optica}. 
Differences in the rates of photorefraction and relaxation would therefore be expected. 
In subsequent experimental studies, the relative effects of the surfaces and bulk can be investigated by changing the width and, consequently, the surface-to-volume ratio of the device, which should, in principle, alter the decay components. 
Furthermore, increasing the dimensions of the nanocavity can alter the ratio of negatively to neutrally charged NV centres, as observed in nanodiamonds~\cite{Rondin2010}. 
Alternatively, surface termination~\cite{Kaviani2014} or surface passivation~\cite{Salvestrini2011,Guha2017,Holzgrafe2024} can alter the surface charge distribution. 
For example, hydrogen termination of the diamond surface has been shown to stabilise the positive and neutral charge state of NV~\cite{Hauf2011} and SiV~\cite{Zhang2023PRL} centres in diamond, respectively. While offering potential insights, a systematic study of the photorefractive effect and relaxation with different surface termination is beyond the scope of this work.

\subsection{Importance of green light for photorefraction}
\begin{figure}
\centering
    \includegraphics[width = \columnwidth]{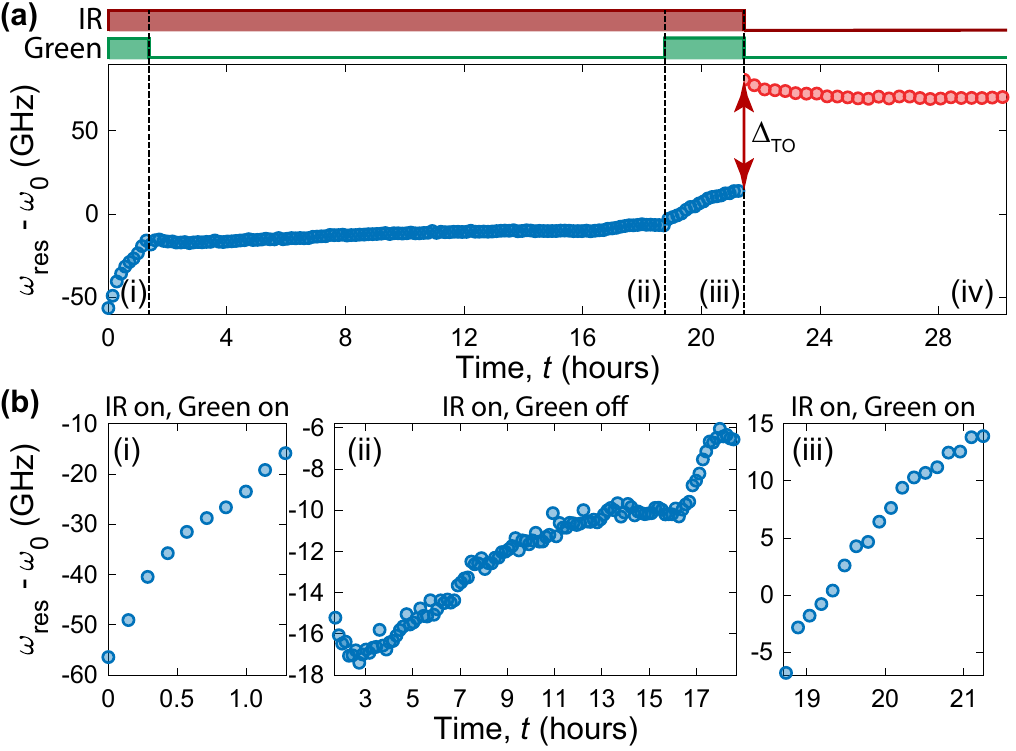}
    \caption{The significance of green light on blue-shifting of the resonance frequency. 
    \textbf{(a)} Acceleration of the blue-shifting is observed when the green and IR lasers are simultaneously injected through the fibre-taper (region (i) and region (iii)).    
    Switching the green laser off results in a drastically reduced rate of blue-shift (region (ii)). 
    Finally, in region (iv) relaxation is observed in the absence of laser illumination.
    The large spectral jump ($\Delta_{\textrm{TO}}\sim11\,\text{GHz}$) between regions (iii) and (iv) can be explained by the lack of red-shift due to thermo-optic effects when the IR laser is switched off. 
    \textbf{(b)} Zooming in on the relevant regions of (a) showcases the accelerated blue-shift when the green laser is on. Note the different scaling on the $x$- and $y$-axes.
    }
    \label{fig:Fig_5}
\end{figure}

Finally, we investigate the importance of the green light on the photorefractive blue-shifting. 
To this end, we perform a control measurement where we simultaneously inject a green laser ($\lambda=520\,\text{nm}$, $P\sim3\,\textrm{mW}$) and the IR laser at high power into the fibre-taper and measure the cavity mode frequency as a function of time (\Cref{fig:Fig_5}).
The green evanescent field of the fibre-taper scatters into waveguide modes of the device. As alluded to above, these modes may be leaky as the device is not designed to support any confined cavity modes at visible wavelengths. 
We observe that the blue-shifting accelerates when the green laser is applied, as can be seen in region (i) and (iii) in \Cref{fig:Fig_5}\,(a,b).
The difference in the rate of blue-shifting between \Cref{fig:PR_effect}\,(a) and \Cref{fig:Fig_5} is attributed to the difference in green intracavity intensity.
For the control measurement, the $\sim3\,\textrm{mW}$ of green laser power injected into the fibre-taper likely exceeds the intensity of the green light generated via THG.
Considering that the photoionisation rates are power-dependent\,\cite{Shandilya2024arxiv}, the increased green intensity will liberate more charges, leading to a faster modulation of the refractive index, and consequently accelerating the blue-shift.
This is consistent with the slower blue-shift observed in the absence of the green laser (\Cref{fig:Fig_5}\,(b-ii)). Reintroducing the green laser in region (iii) again accelerates the blue-shifting.
For completeness, we probe the cavity resonance frequency in without both the IR and green lasers. 
We observe that the cavity resonance frequency jumps by $\Delta_{\text{TO}}\sim 11\,\text{GHz}$, which can be explained by the absence of thermo-optic red-shift and is consistent with the frequency jump observed for $t=0$ in \Cref{fig:PR_effect}\,(b). 
Therefore, based on this control experiment, we conclude that green light is paramount to the observation of photorefraction in diamond. 

\section{Outlook: Applications and Future Directions}
\label{sec:applications}
\subsection{Green light generation in diamond cavities}
The generation of green light is a cornerstone for a myriad of photonic applications\,\cite{Wang2026LSA}. As there are no compact and cost effective solid-state lasers, green light is typically generated by nonlinear frequency conversion processes\,\cite{Corato-Zanarella2025}. 
This work is the first demonstration of green light generation in a diamond nanocavity. We acknowledge that in this demonstration, the third-harmonic conversion efficiency is limited by the modest $Q$-factor compared to other mature photonic platforms, like silicon nitride\,\cite{Wang2026LSA} or lithium niobate\,\cite{Lin2019PRL}.
However the wide bandgap and excellent thermal properties makes diamond capable of handling large on-chip field intensities with limited absorption or thermal damage. Therefore, despite its weaker nonlinearity, diamond is an attractive material platform for green light generation with large output intensities.

Furthermore, as alluded to above, this work is the first demonstration of integrated nonlinear optics in the visible regime in diamond, and facilitates the study of $\chi^{(3)}$ processes in diamond.
Of particular interest is a deeper understanding of the relationship between $\chi^{(3)}$ and the density of crystal defects---as demonstrated by Talik et. al in Ref.\,\cite{Talik2025}, the magnitude of $\chi^{(3)}$ scales inversely with NV centre density.
Therefore, to maximise the THG conversion efficiency, it is desirable to operate with ultrapure diamond material\,\cite{Abulikemu2023}. Nevertheless, third-harmonic generation in NV-centre-rich diamond offers interesting applications.
Beside from the emergence of the photorefractive effect, which we investigate extensively in this work, the internally generated green light can excite photoluminescence from NV centres, which opens the door for IR laser-driven optically detected magnetic resonance (ODMR) protocols. Therefore, third-harmonic generation in NV centre rich diamond paves the way for the realisation of a compact on-chip IR excited NV centre magnetometers.

\subsection{Photorefractive resonance tuning}
Nonvolatile tuning of cavity resonances by the photorefractive effect is a powerful tool for applications in quantum- and nonlinear optics. 
For example, the efficiency of spin-photon interfaces based on diamond colour centres is limited by the photon collection efficiency~\cite{Flagan2022}, which can be greatly enhanced by the Purcell effect~\cite{Purcell1946} when high ${Q}/ {V_{\text{eff}}}$ optical resonances\,\cite{Faraon2011,Barclay2011,Riedel2017,Bhaskar2020,Zifkin2024,Herrmann2024} are spatially and spectrally aligned with the colour centre dipole moment~\cite{Riedel2020}. 
While spectral overlap can be adjusted \textit{in situ} by tuning the emitter using strain~\cite{Meesala2018,MacHielse2019,Brevoord2025}, electric-~\cite{Tamarat2006,Acosta2012,Schmidgall2018,Aghaeimeibodi2021,DeSantis2021}, or magnetic~\cite{Gritsch2023} fields, for example, tuning the cavity resonance frequency is desirable for many applications. 
The most common approaches used to tune a cavity rely on either heating\,\cite{Nitiss2023} or gas condensation~\cite{Faraon2011, Evans2018,Lukin2020NatPhot,Rugar2021}, both of which red-shift the cavity resonance frequency. 
In contrast, the effect studied here allows blue tuning of the cavity resonance, and was not observed to degrade the cavity $Q$.
Furthermore, the photorefractive effect demonstrated in this work allows for independent tuning and control of individual cavities, whereas gas condensation and temperature tuning will affect all devices on the same chip.
Precise frequency trimming of individual cavities post-fabrication has been demonstrated using different methods, such as ultraviolet-\,\cite{Savchenkov2003,DePaoli2020,Farmakidis2023} and electron beam exposure\,\cite{Thiel2022} and AFM assisted nano-oxidation\,\cite{Hatipoglu2022}.
These techniques can achieve $\sim\textrm{nm}$ tuning range at the expense of permanent modifications to the cavity and substantial experimental overhead.
Furthermore, thermally activated, long-range and bi-directional frequency trimming has been recently demonstrated in silicon photonics\,\cite{Belogolovskii2025,Xue2025OpticsExpress}.
However, these techniques require illumination with multiple lasers, and therefore comes at the cost of added experimental complexity. In our work, we demonstrate frequency tuning using a single IR laser only.
Therefore, in the context of diamond photonics, the photorefractive effect constitutes a novel tool to tune and overlap the resonance frequency of several individual devices fabricated on the same chip, which is advantageous for on-chip quantum photonics\,\cite{Katiyi2025}, quantum information processing and metrology\,\cite{Luo2023LightSciApp,Labonte2024} using photonic integrated circuits\,\cite{Wan2020,Kim2020Optica,Wang2020NatPhotRev}.
The slow relaxation process ensures that the colour centre remains resonant with the cavity mode over long timescales. 
Note that while the NV concentration in the sample studied here is higher than that in ultrapure samples used for most experiments with single colour centres, recent measurements have identified optically coherent NV centres in similar diamond material, thereby demonstrating the potential of this material for quantum applications~\cite{Orphal-Kobin2023}.

Diamond material with a high density of NV centres constitute a leading platform for quantum sensing\,\cite{Barry2020,Sorensen2025arXiv},
and the photorefractive cavity tuning demonstrated here could push this technology to achieve higher sensitivity levels. 
Specifically, the NV centre exhibits spin-state-dependent absorption of $1042\,\text{nm}$ light, which can be correlated to changes in environmental parameters like magnetic field\,\cite{Doherty2013}.  
The effectiveness of absorption-based magnetometry schemes is limited by the inherently small IR absorption cross-section of the NV centre; however, resonant recirculation of light in a cavity can greatly increase the absorption probability\,\cite{Jensen2014,Schall2025}. 
Therefore, the photorefractive blue shifting demonstrated in this work can be utilised to tune a cavity onto resonance with the $1042\,\text{nm}$ singlet transition, thereby paving the way for on-chip efficient absorption-based diamond magnetometers\,\cite{Younesi2025}.
Furthermore, independent photorefractive tuning of multiple devices fabricated on the same chip could pave the way for the realisation of dense arrays of fibre-coupled magnetic field sensors.

The \textit{in situ} blue-tuning of the resonant cavity frequency will also be beneficial for realising doubly-resonant cavities for applications in nonlinear optics such as four wave mixing, Brillouin scattering, and Raman scattering. Enhanced nonlinear optical interactions have been demonstrated using optical double-resonances in monolithic structures across various material platforms, such as racetrack resonators~\cite{Latawiec2015,Chen2019Optica,Clementi2025}, microdisks~\cite{Lake2016,Wang2021LSA,Flagan2025}, microrings~\cite{Leo2016,Surya2018,Lukin2020NatPhot,Wang2022NatCom}, and photonic crystal cavities~\cite{Rivoire2010,Minkov2019}.

To better understand the potential of photorefractive tuning, additional studies are needed.
The proposed mechanism underlying this effect---modification of the local charge environment via ionisation of NV centres---can be probed through photoluminescence measurements of the NV charge state during the tuning process.
Similarly, studies on samples with higher NV concentrations are of interest\,\cite{Sorensen2025arXiv,Coutts2026arXiv}, as they may allow larger tuning ranges.
The paramount importance of the THG-generated green light to the modification of the cavity properties
was asserted by simultaneously injecting green and IR lasers into the device (Fig.\,\ref{fig:Fig_5}). Therefore, improving the THG conversion efficiency by tailored device engineering\,\cite{Rivoire2011,Buckley2014,Minkov2019}, would enable a larger and more rapid spectral shift. Furthermore, wavelength dependent excitation\,\cite{Flagan2025} may provide valuable insight into the dominant ionisation processes required for the formation of $E_{\text{sp}}$.
Finally, investigation of techniques for more rapidly and completely resetting the cavity resonance frequency, i.e., by heating or thermal annealing\,\cite{Kavatamane2026} of the sample, is required.
Alternatively, illuminating the device with more energetic light might liberate charges trapped at deep lying defects, such as the neutral divacancy\,\cite{Pu2001,Deak2014,Gorlitz2022}.
In principle, thermally driven relaxation can be performed in tandem with high-power laser illumination, thus providing a potential method to accelerate the recovery of the initial cavity resonance frequency.

\section{Conclusion}
\label{sec:Conclusion}
In conclusion, we have demonstrated cavity-enhanced THG from a diamond photonic crystal cavity, with an end-to-end conversion efficiency of ${\tilde{\eta}_\textrm{THG}=6\cdot10^{-7}\,\text{W}^{-2}}$. Further, we show simultaneous thermo-optic and photorefractive effects, with the latter leading to a blue-shifting of the cavity resonance by ${\Delta\omega /  2\pi= 20.2\,(2)\,\text{GHz}}$, exceeding the cold-cavity linewidth. Photorefractive relaxation occurred over several tens of hours, and both processes are well-described by an analytical framework. The combination of thermo-optic and photorefractive effects offers bi-directional tuning of the cavity resonance, potentially allowing the cavity to be tuned onto resonance with colour centres or the establishment of double-resonance conditions to enhance the efficiency of nonlinear optical interactions.
Furthermore, the photorefractive effect constitutes a novel tuning mechanism for diamond photonics. Contrary to heating or gas condensation, which affect all the devices on the entire chip, photorefraction enables independent tuning of individual devices, offering an invaluable addition to integrated photonic quantum devices in diamond.

\section*{Acknowledgment} We sincerely thank Kartik Srinivasan, Tim Schr{\"{o}}der and Gregor Pieplow for fruitful discussions.
This work was supported by NSERC (Discovery Grant program), Alberta Innovates, and the Canadian Foundation for Innovation. SF acknowledges support from the Swiss National Science Foundation (Project No. P500PT\_206919).


%

\end{document}


\title{Photorefractive tuning seeded by third-harmonic light in a diamond photonic crystal cavity: supplemental document}
\author{} 
\begin{abstract}
\end{abstract}

\maketitle
\section{Experimental Methodology}\label{sec:setup}
\begin{figure}[b!]
    \centering
    \includegraphics[width=\columnwidth]{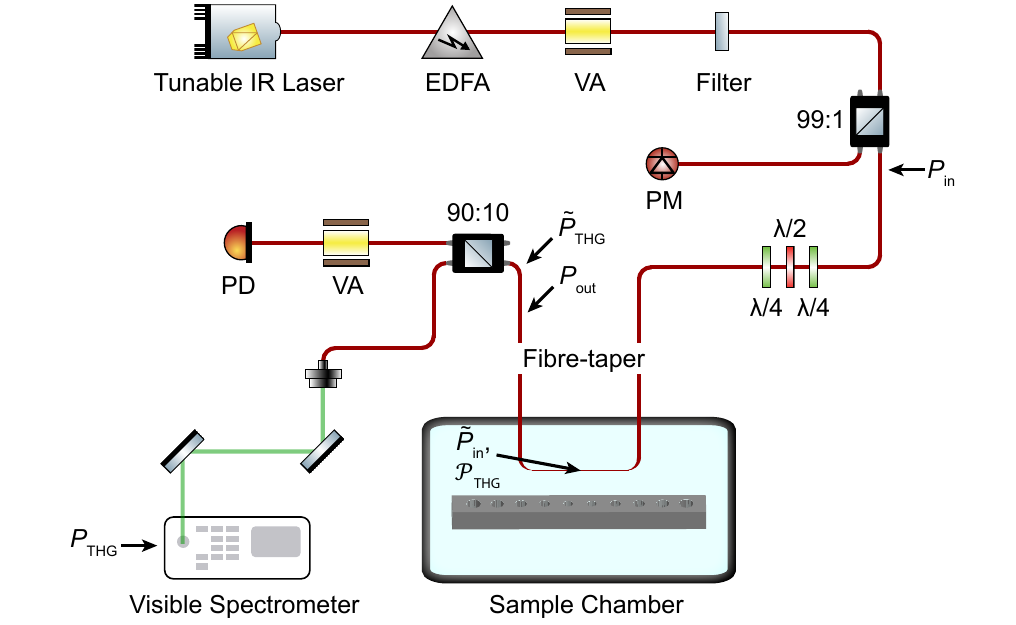}
    \caption{Schematic of the experimental setup used in this work. Read the main text for a discussion of the various components. The labels $P_{\text{in}}$ and $\tilde{P}_{\text{in}}$ respectively correspond to the power measured at the input of the fibre-taper and the power estimated at the tapered region, based on the fibre-taper transmission. 
    Similarly, $P_{\text{THG}}$ is the THG power measured at the spectrometer, while $\tilde{P}_{\text{THG}}$ and $\mathcal{P}_{\text{THG}}$ are the THG powers estimated at the output of the fibre-taper and in the fibre-taper immediately following the cavity, respectively.    
    }
    \label{fig:Experiment_setup}
\end{figure}
The experimental setup used in this work is sketched in \Cref{fig:Experiment_setup}.
The output of a widely tunable continuous wave laser (Santec TSL-710, $\lambda=1480\sim1640\,\text{nm}$) is amplified by an erbium-doped fibre amplifier (EDFA, Pritel LNHPFA-30), and the optical power injected into the cavity is controlled using a variable attenuator (VA, EXFO FVA-3100) placed after the EDFA. The fibre amplifier introduces undesired noise, which we reduce using a wavelength demultiplexer (WDM, MFT-MC-55-20-AFC/AFC-1). We next use a 99:1 fibre beamsplitter (Thorlabs TW1550R1A1), which directs $1\,\%$ of the light to a powermeter (Thorlabs PM400) to monitor the input power. The remaining light is injected into the fibre-taper waveguide. 

\begin{figure}[t!]
    \centering
    \includegraphics[width=\columnwidth]{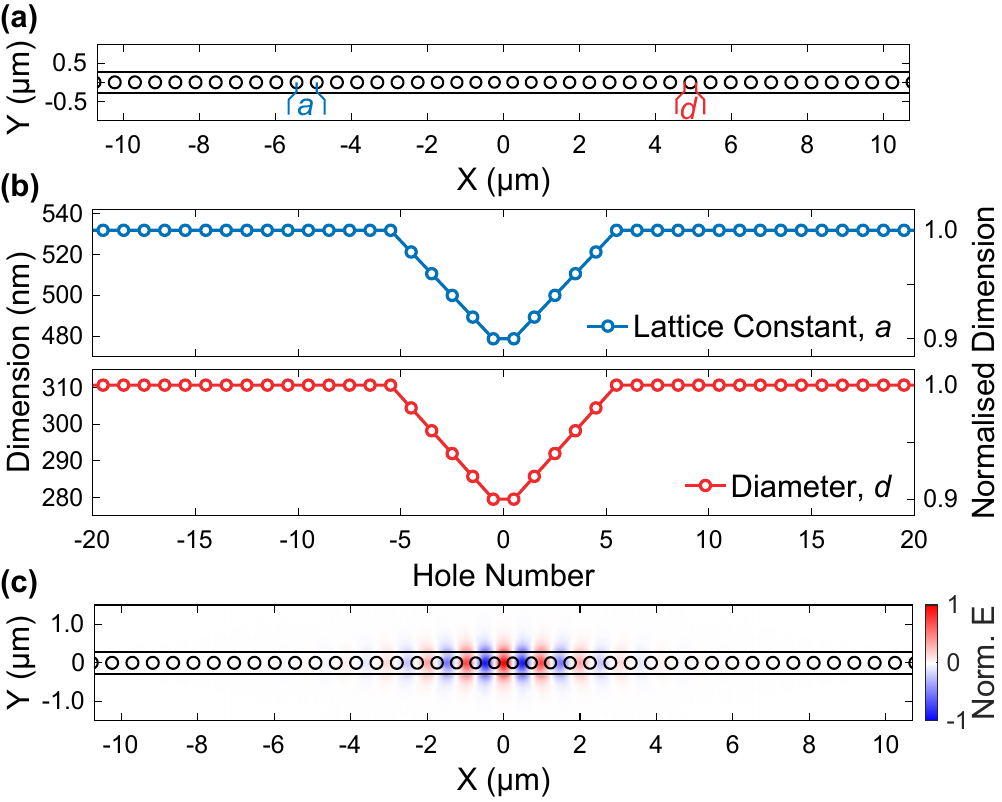}
    \caption{The design of the photonic crystal cavity. \textbf{(a)} The PCC consists of a nanobeam with an array of holes.
    \textbf{(b)} The cavity is formed by introducing a gradual defect in the otherwise periodic array of air holes perforating the waveguide. The defect is created by varying the hole spacing ($a$) and diameter ($d$) as shown in the image.
    \textbf{(c)} The distribution of the normalised in-plane electric field ($E_\text{y}$) of the cavity mode at $\lambda_\text{0} \approx 1565\, \text{nm}$ predicted from simulations.
    }
    \label{fig:Comsol}
\end{figure}

Central to the experiment is a photonic crystal cavity (PCC), fabricated from `optical grade' bulk single-crystal diamond (Element Six, [N$_{\text{s}}$]$\sim1\,\text{ppm}$ and [NV]$\sim0.01\,\text{ppm}$) using the quasi-isotropic undercut method. Details of this fabrication procedure can be found elsewhere\,\cite{Khanaliloo2015, Khanaliloo2015NanoLett,Mitchell2016, Mouradian2017Rectangular, Mitchell2019}. 
The PCC consists of a suspended waveguide with width $w\sim 550\,\text{nm}$ and length $l\sim30\,\upmu\text{m}$, with an array of holes patterned in the centre of the waveguide, as shown schematically in \Cref{fig:Comsol}\,(a).
Following design principles presented elsewhere \cite{Chan2009,Quan2011OptExp}, the diameter of the holes, $d$, and the hole spacing or lattice constant, $a$, are symmetrically tapered as shown in \Cref{fig:Comsol}\,(b) to create an optical cavity in the centre of the waveguide. The nominal value of $a$ in the mirror regions on each side of the cavity defect was chosen to create an optical bandgap in the 1550 nm wavelength range. The optical modes of the resulting cavity were simulated using COMSOL Multiphysics finite element solver, and the electric field distribution of the cavity's fundamental TE-like mode is shown in \Cref{fig:Comsol}\,(c). 

As described below, light resonant with the cavity mode is coupled into and out of the PCC evanescently.  Third-harmonic-generated light is collected by the fibre-taper via the same coupling mechanism. To accomplish this, we position the fibre-taper within the near-field of the PCC using stepper-motor translation stages (Suruga Seiki XXC06020-G)\,\cite{Masuda2024}.
Typically the separation between the fibre-taper and the photonic crystal cavity is on the order of $\sim300-500\,\text{nm}$. The length of the coupling region between the fibre-taper and the photonic crystal is given by the spatial extent of the cavity mode. From the simulations in \Cref{fig:Comsol}\,(c), the cavity mode extends for approximately $4\,\upmu\text{m}$.
A polarisation paddle controller is used to vary the polarisation of the incoming light to maximise the coupling to the PCC.
To maintain a constant separation and corresponding coupling between the fibre-taper and the device, we park the fibre on the support beams next to the cavity, as shown in Fig.\,1\,(a) of the main manuscript. To avoid undesired contact between the fibre-taper and the diamond substrate, the fibre-taper is fabricated with a \textit{dimple}\,\cite{Michael2007,Masuda2024}. Transmission through the fibre-taper was characterised using the IR laser and a $532\,\text{nm}$ CW laser and measured to be $\eta_{\text{fibre}}(1550\,\text{nm})=23\,\%$ and $\tilde{\eta}_{\text{fibre}}(532\,\text{nm})=15\,\%$, respectively. The $532\,\text{nm}$ laser is the closest available laser source to the third-harmonic-generated light at $\sim522\,\text{nm}$ --  we assume that the fibre-taper transmission remains constant across this $10\,\text{nm}$ difference in wavelength. 

\begin{table*}[bth!]  
\centering 
\caption{Summary of definitions for the powers and efficiencies used in this work.} 
\begin{tabularx}{1\textwidth}{l@{\extracolsep{\fill}}  l } 
\hline\hline   
Parameter  & Description\\ 
\hline\hline  
$\eta_\textrm{fibre} = 0.23$ & Measured fibre-taper transmission at $1566\,\textrm{nm}$ \\
$\tilde{\eta}_\textrm{fibre} = 0.15$ &   Measured fibre-taper transmission at $532\,\textrm{nm}$\\
$P_\textrm{in}$ & Measured IR laser power at the input of the fibre-taper \\
$\tilde{P}_\textrm{in} = \sqrt{\eta_\textrm{fibre}}P_\textrm{in}$ & Inferred IR power at tapered region of the fibre-taper  \\
$\tilde{\eta}_\textrm{link}$ = 0.0035 & \makecell[l]{Transmission of the fibre link between the output of the fibre-taper and the spectrometer measured at $532\,\textrm{nm}$}\\
$P_\textrm{THG}$ & Measured THG power at the spectrometer\\
$\tilde{P}_\textrm{THG} = P_\textrm{THG}/\eta_\textrm{link}$ & Inferred THG power at the output of the fibre-taper\\
$\mathcal{P}_{\text{THG}}=\tilde{P}_{\text{THG}}/\sqrt{\tilde{\eta}_{\text{fibre}}}$ & THG power in the fibre-taper immediately following the cavity\\
\hline 
\hline
\end{tabularx} 
\label{table:summary_notation}
\end{table*}

The output of the fibre-taper is injected into a 90:10 fibre beamsplitter (Newport F-CPL-L22151-A). 
The 10\,\% output port is connected to a photodiode (PD, Newport 1623) used to monitor the fibre-taper transmission. We use a second variable attenuator immediately before the photodiode to ensure a constant power level at the photodiode for any given input power. Using this configuration, we can perform both high-and low-power laser scans without saturating the photodiode or changing the electronic gain of the photodiode. We direct the remaining 90\,\% of the transmitted light to a free-space spectrometer (Princeton Instruments Acton SP2750 with PIXIS 100B CCD detector) used to monitor the intensity of the THG signal. Before the experiment, a green CW laser ($532\,\text{nm}$) was used to calibrate the spectrometer efficiency, enabling us to convert detected CCD counts into THG power.

To achieve accurate fibre-taper transmission measurements, it was crucial to determine the noise-floor voltage value of the photodiode. To that end, we set the attenuation of both attenuators to the maximum value of $\sim 70\,\text{dB}$ (see \Cref{fig:Experiment_setup}) and recorded the fibre-taper transmission as a function of wavelength. For all of the fibre-taper transmission data shown in the main text, we subtract the noise-floor voltage value before normalising the fibre-taper transmission.

To properly interpret the analysis, it is important to distinguish between the power injected into the fibre-taper and the power at the tapered region of the fibre. We therefore adopt the following notation: the power injected into the fibre-taper is denoted as $P_{\text{in}}$ and is inferred from the splitting ratio of the 99:1 fibre beamsplitter and the power measured at the powermeter.
The power at the fibre-taper cavity interface can be estimated from the measured fibre-taper transmission efficiency $\eta_{\text{fibre}}$ by assuming that the fibre-taper is equally lossy on either side of where it interacts with the cavity. In other words, the power in the fibre-taper at the cavity is given by $\tilde{P}_{\text{in}}=\sqrt{\eta_{\text{fibre}}}P_{\text{in}}$.
We adopt a similar notation for the detected third-harmonic-generated light, with $P_{\text{THG}}$ being the THG power detected at the spectrometer and $\tilde{P}_{\text{THG}}$ being the inferred power at the output of the fibre-taper.
When calculating $\tilde{P}_{\text{THG}}$, we account for the fibre-link efficiency of $\tilde{\eta}_{\text{link}}= 0.0035$ measured using the 532\,nm laser. In \Cref{fig:Experiment_setup}, we highlight where the different powers are measured and inferred. The fibre-link efficiency, $\tilde{\eta}_{\text{link}}$, is extremely low because of the optical fibre used and the many fibre connections required to connect the output of the fibre-taper to the spectrometer.
The fibre connection between the fibre-taper, designed to be single mode at IR wavelength, and the visible single-mode fibre leading to the spectrometer is substantially lossy. 
Finally, we define $\mathcal{P}_{\text{THG}}=\tilde{P}_{\text{THG}}/\sqrt{\tilde{\eta}_{\text{fibre}}}$ to be the THG power in the fibre-taper immediately following the cavity.
In Table.\,\ref{table:summary_notation} we summarise the definition and value of the various power and efficiencies used in this work.

In the experimental configuration, the fibre-taper waveguide is the most power sensitive component. For example, increased absorption by a dust particle can lead to an increased temperature and consequential thermal damage to the fibre-taper. 
All the measurements were performed at ambient condition, where convection is the dominant source of heat dissipation. We therefore limit the maximum laser power injected into the fibre-taper waveguide to $\sim75\,\text{mW}$ to avoid thermal damage.
It is also worth noting  that increasing the laser power will eventually lead to the cavity becoming thermally bistable\,\cite{Carmon2004,Shankar2011}.

\section{Calculating the Effective Mode Volume}
Here, we calculate the effective mode volume of the cavity mode, which allows the number of photons circulating in the cavity to be converted to peak field intensity. To start, we calculate the electric field amplitude per photon\,\cite{Barclay2007}. In general, the electric field is quantised according to\,\cite{Flagan2022}
\begin{equation}
    \int_{V}\epsilon(\bm{r}')\left|\bm{E}(\bm{r}')\right|^2d^3\bm{r}'=\frac{\hbar\omega_{\text{cav}}}{2}\,,
\end{equation}
from which the vacuum electric field amplitude is given by\,\cite{Khitrova2006,Riedel2017}
\begin{equation}
    \bm{E}(\bm{r})=\sqrt{\frac{\hbar\omega_{\text{cav}}}{2\epsilon(\bm{r})V_r(\bm{r})}}\,.
\end{equation}
Here, $\epsilon(\bm{r})$ is the permittivity of the dielectric medium defining the cavity and 
\begin{equation}
    V_r(\bm{r})=\frac{\int\epsilon(\bm{r}')\left|\bm{E}(\bm{r}')\right|^2d^3\bm{r}'}{\epsilon(\bm{r})\left|\bm{E}(\bm{r})\right|^2}
\end{equation}
is a generalised mode volume at position $\bm{r}$\,\cite{Barclay2007,Santori2010}. 

A commonly used figure of merit is the optical mode volume $V_{\text{o}}$, defined from the peak energy density according to\,\cite{Barclay2007}
\begin{equation}
    V_{\text{eff}}=\frac{\int\epsilon(\bm{r}')\left|\bm{E}(\bm{r}')\right|^2d^3\bm{r}'}{\text{max}\left[\epsilon(\bm{r}')\left|\bm{E}(\bm{r}')\right|^2\right]}\,,
    \label{eq:modeVolumeEff}
\end{equation}
from which we can define a dimensionless mode volume
\begin{equation}
V_{\text{o}}=V_{\text{eff}}/\left(\frac{\lambda_{\text{cav}}}{n_0}\right)^{3}\, ,
\end{equation}
where $n_0$ is the refractive index of diamond.

To calculate the effective mode volume, we simulate the electric field profile of the cavity mode using COMSOL Multiphysics finite element solver. From the fundamental IR mode shown in \Cref{fig:Comsol}\,(c), we calculate $V_{\text{eff}}=0.13\,\times\left(\frac{\lambda_{\text{cav}}}{n_0}\right)^{3}$. From the experimentally measured cold-cavity $Q$-factor of $\sim10^{4}$ (Fig.\,1\,(b) of the main manuscript), we determine $\frac{Q}{V_{\text{eff}}}\sim10^{5}\times\left(\frac{\lambda_{\text{cav}}}{n_0}\right)^{-3}$.
We next calculate the confinement factor, which describes the fraction of the mode's energy that is confined within the diamond regions of the cavity\,\cite{Barclay2007}
\begin{equation}
\Gamma_{\text{o}}=\frac{\int_{\text{diamond}}\epsilon(\bm{r}')\left|\bm{E}(\bm{r}')\right|^2d^3\bm{r}'}{\int\epsilon(\bm{r}')\left|\bm{E}(\bm{r}')\right|^2d^3\bm{r}'}\,.
\end{equation}
From our simulated mode profile we find $\Gamma_{\text{o}}=0.84$.

\section{Calculating the peak cavity intensity}
The peak intensity inside the cavity is given by~\cite{Hecht2016}
\begin{align}
    I_{\text{peak}} = \frac{1}{2}\frac{c}{n_\text{g}}\epsilon\left|\bm{E}_\text{max}\right|^2\,,
    \label{eq:peakIntensity}
\end{align}
where $\bm{E}_{\text{max}}$ is the maximum electric field vector inside the cavity, $\epsilon$ is the dielectric permittivity at the position of $\bm{E}_\text{max}$, $c$ is the speed of light in vacuum, and $n_\text{g}$ is the group index of the optical cavity mode. 
Therefore, to find the peak intensity, we simply need to determine the group index, the maximum field amplitude, and its corresponding permittivity.

First, we use the phase velocity of the photonic crystal waveguide to approximate the group index:
\begin{align}
    n_\text{g} = c\frac{k}{\omega_{\text{cav}}}\,.
    \label{eq:group_index}
\end{align}
Here, $k = \pi/a$, where $a=480\,$nm is the waveguide lattice constant at the position of the cavity mode (see \Cref{fig:Comsol}\,(a)). 
These values give a group index of $n_\text{g}=1.63$.
Next, we find the maximum field amplitude, which, as shown in \Cref{fig:Comsol}\,(c), is located within the diamond. Thus, $\epsilon$ is the permittivity of diamond at the cavity resonance frequency~\cite{Phillip1964}.
Lastly, $|\bm{E}_{\text{max}}|$ can be calculated using the following equation~\cite{Khitrova2006,Flagan2022}
\begin{align}
    |\bm{E}_{\text{max}}| = \sqrt{\frac{n_\text{cav}\hbar \omega_\text{cav}}{2\epsilon V_\text{eff}}}\,,
    \label{eq:maxField}
\end{align}
where $V_{\text{eff}}$ is the effective mode volume introduced in \Cref{eq:modeVolumeEff}, and $n_{\text{cav}}$ is the intracavity photon number. 
Combining \Cref{eq:peakIntensity,eq:maxField}, we find a new expression for the peak intensity:
\begin{align}
    I_\text{peak} = \frac{1}{4} \frac{n_\text{cav}c\hbar \omega_\text{cav}}{n_\text{g} V_\text{eff}}\,.
\end{align}
Using the experimental parameters summarised in Tab.\,1 of the main text and in the previous section, we find the peak intensity to be $I_\text{peak}=56\,\text{GW}/\text{cm}^2$, which corresponds to loading the cavity with $3.5\times10^6$ photons (see \Cref{sec:Thermo-optic_effects}).

\section{Coupled mode theory}
\begin{figure}[t!]
    \centering
    \includegraphics[width=\columnwidth]{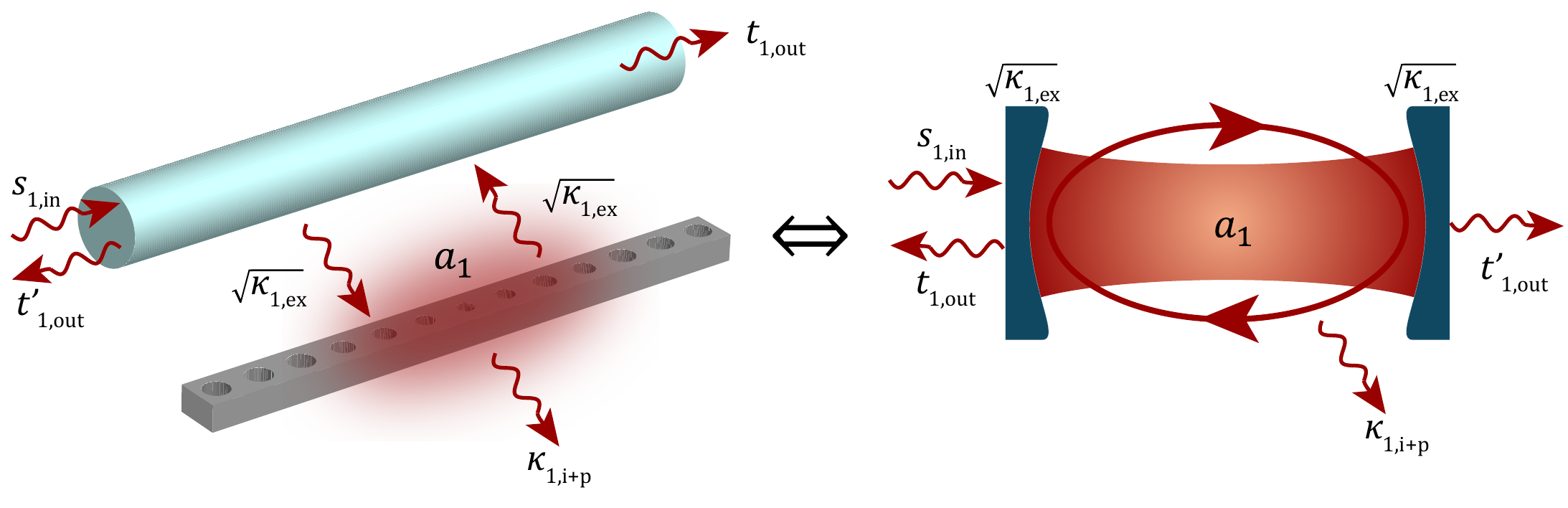}
    \caption{The coupling between a fibre-taper waveguide mode and a PCC (left) is equivalent to a canonical Fabry-Perot cavity (right).} 
    \label{fig:PhX_Fabry_Perot}
\end{figure}

Coupled mode theory provides the formalism necessary to describe the coupling between the evanescent field of a fibre-taper waveguide and an optical mode in a photonic resonator\,\cite{Aspelmeyer2014,Sun2017OpticsExpress,McLaughlin2022}. By employing coupled mode theory, we can relate the optical properties of the cavity, such as the intracavity intensity and its loss-rates, to the fibre-taper transmission spectrum. For simplicity, we assume a mean-field approximation, where the number of laser pump photons greatly exceeds the variations in photon number due to quantum fluctuations. In a frame of reference co-rotating with the laser frequency, the equation of motion describing the amplitude of the cavity mode in a fibre-taper waveguide coupled cavity is\,\cite{Aspelmeyer2014,McLaughlin2022}
\begin{equation}
    \dot{a}_1 = \left(i\Delta_0 - \frac{\kappa_1}{2}\right)a_1 + \sqrt{\kappa_{\text{1,ex}}}\;s_\text{1,in}\,,
    \label{eq:CME_1}
\end{equation}
where $\langle\hat{a}_{1}\rangle = a_1$ is the electric field amplitude of the cavity mode and $\langle\hat{s}_\text{1,in}\rangle= s_\text{1,in}=\sqrt{\tilde{P}_{\text{in}}/\hbar\omega}$ is the electric field amplitude of the input laser. The term $\Delta_0=\omega-\omega_0$ describes the frequency detuning between the input laser field and the cavity resonance, at frequency $\omega$ and $\omega_0$, respectively, while $\kappa_{1}$ is the total energy decay rate of the cavity. 
The extrinsic decay rate, $\kappa_\text{1,ex}$, is the coupling rate between the fibre-taper input field and the cavity mode and is related through unitarity to the optical coupling coefficient $\sqrt{\kappa_\text{1,ex}}$.
Further, because the PCC mode is a standing wave, it couples equally well to forward and backward propagating fields in the fibre-taper. This introduces two extrinsic loss channels, allowing us to express the total decay rate as $\kappa_1 = \kappa_\text{1,i+p} + 2\kappa_\text{1,ex}$, where $\kappa_\text{1,i+p}$ encompasses the intrinsic cavity loss of the cavity and any additional parasitic loss introduced by the fibre-taper interaction such as coupling to non-fundamental fibre-taper modes or scattering into radiation modes \cite{Spillane2003}.
In \Cref{fig:PhX_Fabry_Perot}, we depict the coupling between the fibre-taper and the PCC, highlighting its equivalence to a canonical Fabry-Perot cavity.

For the device studied here, the time required to load the cavity to its steady-state condition is on the order of $10^{-10}\,\text{s}$, as determined by the cavity linewidth (see main manuscript). As this time-scale is much faster than any experimental time-scales, a steady-state approximation is valid. In the steady-state approximation, where $\dot{a}_1 = 0$, the amplitude of the optical cavity field is given by
\begin{equation}
    \label{eq:cold_a}
    a_1 = \frac{\sqrt{\kappa_{\text{1,ex}}}}{-i\Delta_0 + \frac{\kappa_1}{2}}\;s_\text{1,in}\,.
\end{equation}

The fibre-taper transmission is determined from input-output formalism\,\cite{Aspelmeyer2014,Clerk2010PhysRev,Mitchell2019UC,McLaughlin2022} as $t_\text{1,out} = s_\text{1,in}e^{i\phi_1} - \sqrt{\kappa_\text{1,ex}}a_1$, where the transmission phase $\phi_1$ accounts for the presence of Fano interference effects\,\cite{Fan2003}.
Finally, by normalising $t_{\text{out}}$ to the input laser field $s_\text{in}$, and taking the complex modulus square, the normalised cavity transmission intensity is given by
\begin{equation}
    \label{eq:transmission_cold}
    T = \left|\frac{t_\text{1,out}}{s_\text{1,in}}\right|^2 = \left|e^{i\phi_1} - \frac{\kappa_{\text{1,ex}}}{-i\Delta_0 + \frac{\kappa_1}{2}}\right|^2\,.
\end{equation}
The photon flux output from the fibre-taper is $|t_{\text{1,out}}|^2$, so that the optical output power of the fibre-taper is given by
\begin{equation}
    \label{eq:THG_power_cold}
    P_{\text{out}} =\sqrt{\eta_{\text{fibre}}}\hbar\omega|t_\text{1,out}|^2 =\sqrt{\eta_{\text{fibre}}}\,  \frac{\Delta_0^2 + \left(\frac{\kappa_1}{2}-\kappa_{\text{1,ex}}\right)^2}{\Delta_0^2 + (\frac{\kappa_1}{2})^2} \tilde{P}_{\,\text{in}}\,.
\end{equation}

\section{Third-harmonic generation}
\begin{figure}[t!]
    \centering
    \includegraphics[width=\columnwidth]{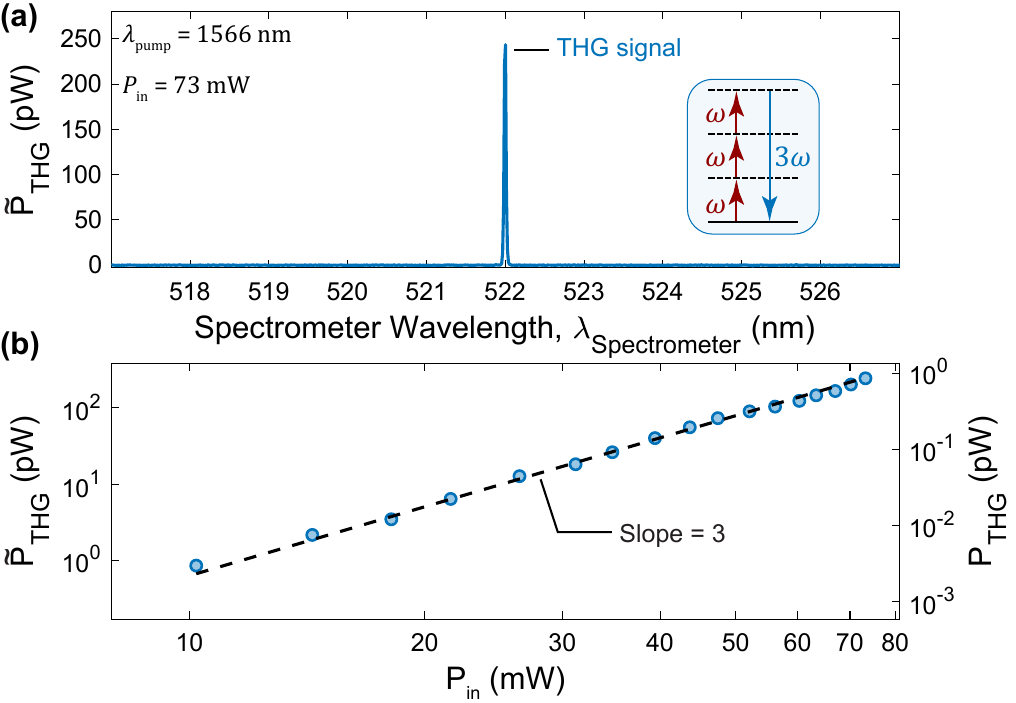}
    \caption{
    \textbf{(a)} Third-harmonic generation from the diamond PCC. In this measurement, the laser was parked on-resonance at $\lambda_{\text{pump}}=1566.01\,\text{nm}$ with $73\,\text{mW}$ injected into the fibre-taper.
    The inset shows the third-harmonic generation process, where three photons at frequency $\omega$ upconvert to one photon at $3\omega$.
    \textbf{(b)} The blue data points show the maximum measured on-resonance THG output power for increasing input power, $P_{\text{in}}$. The right $y$-axis shows the detected THG power on the spectrometer, $P_{\text{THG}}$, while the left $y$-axis shows the estimated THG power at the output of the fibre-taper, $\tilde{P}_{\text{THG}}$.
    The black dashed line shows the cubic power scaling expected for a THG process. 
    }
    \label{fig:THG_Sup}
\end{figure}

When the IR laser was coupled to the cavity at high intensity, we observed the emission of green light. To verify the origin of this photon emission, we perform spectrally resolved measurements with increasing intensity (as shown in Fig.\,2 of the main manuscript). As we show in \Cref{fig:THG_Sup}\,(a), for fixed laser wavelength $\lambda_{\text{pump}}=1566\,\text{nm}$, we observe this photon emission at $\lambda_{\text{pump}}/3=522\,\text{nm}$, which is indicative of third-harmonic generation.
However, to conclusively verify the origin of this photon emission,
we plot the peak on-resonance output power with increasing input power on a log-log scale (\Cref{fig:THG_Sup}\,(b)).
We find that the power of the observed photon emission scales with the input power cubed, i.e. $P_{\text{\,THG}}\propto P_{\text{in}}^3$, where $P_{\text{in}}$ is the measured power injected into the fibre-taper (see \Cref{fig:Experiment_setup}). From this scaling with power, we verify that the visible emission is produced via third-harmonic generation, where $\bm{E}_{3\omega}=\chi^{(3)}\bm{E}_{\omega}^3$\,\cite{Carmon2007}.

To estimate the end-to-end conversion efficiency of the third-harmonic generation process, we first must estimate the losses in the fibre link between the output of the fibre-taper and the slit of the spectrometer. To this end, as described in Sec.\,\ref{sec:setup}, we use a $532\,\text{nm}$ CW laser and measure a link efficiency of  $\tilde{\eta}_{\text{link}}=0.35\,\%$. From the THG signal measured on the spectrometer, $P_{\text{THG}}$, we estimate the THG power exiting the fibre-taper to be $\tilde{P}_{\text{THG}}={P_{\text{THG}}}/{\tilde{\eta}_{\text{link}}}$. The end-to-end conversion efficiency, $\tilde{\eta}_{\text{THG}}$, is then calculated from 
\begin{equation}
    \tilde{\eta}_{\text{THG}} = \frac{\tilde{P}_{\text{THG}}}{P_{\text{in}}^3}\,.
    \label{eq:Estimated_THG_eff}
\end{equation}
From the power dependent measurement shown in Fig.\,2\,(b) of the main text, we estimate $\tilde{\eta}_{\text{THG}}=6\cdot10^{-7}\,\text{W}^{-2}$. 
We note that $\tilde{\eta}_{\text{THG}}$ is strictly the end-to-end efficiency estimated from reliable power measurements performed at the input and output port of the fibre-taper without correcting for the poor coupling of the third-harmonic generated light in the cavity to the guided mode of the fibre-taper.
Therefore, the end-to-end conversion efficiency can be further improved by introducing a separate fibre-taper designed for operation at the THG wavelength\,\cite{Logan2018}.

\subsection{Cavity-Enhanced Third-harmonic Generation}
Here, we will outline the theory describing the enhancement of the third-harmonic generation in the single mode of the PCC. The design of the PCC exhibits no confined modes at or near the third-harmonic frequency -- the PCC will behave like a waveguide at this frequency. However, when modelling our system, we consider the waveguide mode as a lossy cavity defined by backreflections from the waveguide ends, to which the third-harmonic field can couple. The total loss rate of the THG waveguide mode is given by $\kappa_3$, while $\kappa_\text{3,ex}$ is the coupling rate between the cavity and the fundamental fibre-taper mode. From this, the conversion from pump photons in the cavity mode to THG photons in the waveguide mode can be expressed as
\begin{equation}
    \label{eq:CME_THG_pump}
    \dot{a}_{3} = -\frac{\kappa_3}{2}a_3 + i\omega_{3}\;\beta_{\text{THG}}\,a_1^3\,.
\end{equation}
Here, $a_3$ is the optical field amplitude at the THG waveguide mode and $\omega_3$ is the THG photon frequency given by $\omega_3 = 3\omega_0$. The constant $\beta_\text{THG}$, defined as\,\cite{McLaughlin2022}
\begin{equation}
    \label{eq:mode_overlap}
     \beta_{\text{THG}} = \frac{3}{8}\frac{\int \epsilon\chi^{(3)}\left(\bm{E}_1^*\cdot\bm{E}_1^*\right)\left(\bm{E}_1^*\cdot\bm{E}_3\right)d^3\bm{r}}{\left(\int \epsilon\left|\bm{E}_1\right|^2 d^3\bm{r}\right)^{\frac{3}{2}}\left(\int \epsilon \left|\bm{E}_3\right|^2 d^3\bm{r}\right)^{\frac{1}{2}}}\,,
\end{equation}
describes the optical inter-modal coupling and spectral overlap between the pump field, $\bm{E}_1$, and the THG field, $\bm{E}_3$\,\cite{McLaughlin2022}. For simplicity, we assume that nonlinear back conversion ($\omega_3\rightarrow3\omega_0$) is negligible. 

As before, we assume a steady-state approximation, and solve \Cref{eq:CME_THG_pump} to find the THG field amplitude 
\begin{equation}
    \label{eq:THG_a}
      a_3 = i\frac{\omega_{3}}{\left(\frac{\kappa_3}{2}\right)}\beta_{\text{THG}}\;a_1^3\,.
\end{equation}
Next, we utilise input-output formalism for the THG field, where the transmission is expressed as 
\begin{equation}
    t_\text{3,out}=s_\text{3,in}-\sqrt{\kappa_{3,\text{ex}}}a_3\,.
    \label{eq:t_2_out_THG}
\end{equation}
There is no incoming field at the THG wavelength. Therefore, by inserting $s_\text{2,in}=0$ and \Cref{eq:THG_a} into \Cref{eq:t_2_out_THG}, the THG output power is given by
\begin{equation}
    \label{eq:THG_power}
    \begin{split}
    \mathcal{P}_{\text{THG}} &= \hbar\omega_{3}|\sqrt{\kappa_{\text{3,ex}}}a_3|^2 \\
    & =\frac{108}{\hbar^2}\times\mathcal{L}\times\left(\frac{\kappa_{\text{1,ex}}}{\Delta_0^2 + \left(\frac{\kappa_1}{2}\right)^2}\right)^3\tilde{P}_{\text{in}}^3\,,
    \end{split}
\end{equation}
where $\mathcal{L} = \left(\kappa_\text{3,ex}/\kappa_3^2\right)\times|\beta_\text{THG}|^2$.
The conversion efficiency, defined as $P_{\text{THG}}/P_{\text{in}}^3$, can be readily calculated from \Cref{eq:THG_power}. For zero detuning, $\Delta_0=0$, the maximum conversion efficiency is thus given by
\begin{equation}
    \label{eq:THG_conversion_eff}
    \eta_{\text{THG}} =  \frac{108}{\hbar^2}\times\mathcal{L}\times\left(\frac{4\kappa_{1,\text{ex}}}{\kappa_1^2}\right)^3\,.
\end{equation}
We note that we do not have access to a tunable laser centred around the third-harmonic wavelength. We are therefore unable to perform coherent mode spectroscopy to independently measure $\kappa_{3}$ and $\kappa_{3,\text{ex}}$, which is required to accurately quantify both the internal THG intensity and the internal conversion efficiency.
However, we provide an estimate for these parameters in the following section.

\subsection{Estimating the Circulating Third-harmonic Power}
\label{sec:Estimating_green_power}
\begin{figure}[t!]
    \centering
    \includegraphics[width=\columnwidth]{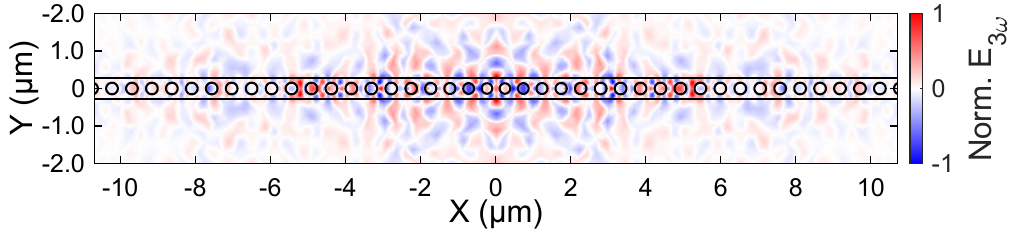}
    \caption{In-plane electric field distribution of the third-harmonic generated light at $522\,\text{nm}$ predicted from simulations. The photonic crystal cavity does not supports any confined modes for this wavelength. From the simulation, we extract $\Gamma_{\text{THG}}=0.2$ and $Q_{\text{THG}}=270$.}
    \label{fig:COMSOL_THG_Field}
\end{figure}

We next attempt to estimate the circulating power of the internally generated green light using the experimentally measured THG power at the the output of the fibre-taper. 
To this end, we start by simulating the electric field distribution at the third-harmonic wavelength using COMSOL Multiphysics to identify potential optical modes.
As can be seen from the example simulation shown in \Cref{fig:COMSOL_THG_Field}, the mode is leaky and not confined to the photonic crystal cavity.
These simulations return several nearly-identical candidate modes, and for the remainder of this calculation, we chose the mode with the largest confinement and quality factors, $\Gamma_{\text{THG}}=0.2$ and $Q_{\text{THG}}=270$, respectively.
From the simulated $Q$-factor, we calculate $\kappa_\text{3,i+p}=\omega_3/Q_{\text{THG}}=2\pi\times2.1\,\text{THz}$.

Starting from \Cref{eq:THG_power}, the number of THG photons in the cavity, $n_3$, can be expressed as 
\begin{equation}
        n_3 = |a_3|^2=\frac{\mathcal{P}_{\text{THG}}}{\hbar \omega_3}\times\frac{1}{\kappa_{3,\text{ex}}}\,.
    \label{eq:n3}
\end{equation}
We next use the intracavity photon number to estimate the circulating power in the cavity
\begin{align}
    P_{\text{circ}}^{\text{THG}} = n_{3}\hbar\omega_3\times \frac{c}{2 n_g L_{\text{eff}}}\,.
    \label{eq:PcircTHG}
\end{align}
Here, $n_{3}\hbar\omega_3$ is the THG energy contained by the cavity and $\tau_{\text{rt}} = 2n_g L_{\text{eff}}/c$ is the cavity round-trip time.
For the THG mode displayed in \Cref{fig:COMSOL_THG_Field}, we approximate $L_{\text{eff}}\approx 10\,\upmu\text{m}$. We calculate $n_g=0.52$ from \Cref{eq:group_index} using $a_{\text{THG}}\sim505\,\text{nm}$ as an approximate value for the lattice constant (see \Cref{fig:Comsol}\,(b)).  
We can therefore estimate the circulating THG power using \Cref{eq:n3,eq:PcircTHG} and the third-harmonic power immediately following the cavity, $\mathcal{P}_{\text{THG}}$, given by
\begin{equation}
\mathcal{P}_{\text{THG}}=\frac{\tilde{P}_\textrm{THG}}{\sqrt{\tilde{\eta}_\textrm{fibre}}}\,,
\end{equation}
which can be estimated from the measurement of the detected THG power and the fibre-taper transmission (see \Cref{table:summary_notation}).
For the highest IR input power, we measure $\tilde{P}_{\text{THG}} \simeq 250\,\text{pW}$ (Fig.\,S4), which corresponds to $\mathcal{P}_{\text{THG}}=0.65\,\text{nW}$. 

As aforementioned, we are unable to experimentally measure $\kappa_{3,\text{ex}}$. Therefore, in order to estimate $P_{\text{circ}}^{\text{THG}}$, we have to make assumptions on $\kappa_{3,\text{ex}}$.
First, as a lower limit, we assume critical coupling, i.e. $\kappa_\text{3,i+p}=2\kappa_{{3,\text{ex}}}$, which yields $P_{\text{circ,\ low}}^{\text{THG}}=2.8\,\text{nW}$.
Next, as an upper limit, we assume that the third-harmonic and IR fields have identical external coupling rates, $\kappa_{{3,\text{ex}}}=\kappa_{{1,\text{ex}}}=2\pi\times6.15\,\text{GHz}$ (see Tab.1 of the main manuscript), which yield $P_{\text{circ,\ high}}^{\text{THG}}=484\,\text{nW}$.

We next use the parameters in Tab.\,1 of the main manuscript, and perform a similar calculation for the circulating IR power.
For maximum IR power, $P_{\text{in}}=73\,\text{mW}$, we find $P_{\text{circ}}^{\text{IR}}=10.2\,\text{W}$. Here, we approximated the effective cavity length to be $L_{\text{eff}}\approx4\,\upmu\text{m}$ (see \Cref{fig:Comsol}\,(c)).

Finally, the finesse $\mathcal{F}$ of the photonic crystal cavity can be calculated using the relation
\begin{equation}
    \frac{P_\text{circ}}{P_0}=\frac{\mathcal{F}}{\pi}\,,
\end{equation}
where $P_0$ is the power dropped into the cavity. 
This calculation yields $\mathcal{F}_\text{IR}=916$ for the IR mode. Similarly, we estimate $\mathcal{F}^{\text{THG}}_{\text{high}}=2358$ and $\mathcal{F}^{\text{THG}}_{\text{low}}=13$ for the upper and lower bound on $P_{\text{circ}}^{\text{THG}}$, respectively.
The comparatively large value of $\mathcal{F}^{\text{THG}}_{\text{high}}$ compared to $\mathcal{F}_\text{IR}$ indicates that the assumption $\kappa_{{3,\text{ex}}}=\kappa_{{1,\text{ex}}}$, i.e. equal coupling rates for the fundamental and third-harmonic fields, likely overestimates $P_{\text{circ}}^{\text{THG}}$.
All the values used in this estimation of circulating green power are summarised in \Cref{table:Estimation_Green_Power}.

\setlength{\extrarowheight}{2pt}
\begin{table}[t]  
\centering 
\caption{Values used to estimate the circulating green power in the device. In this calculation, we have used $\tilde{P}_{\text{THG}} \simeq 250\,\text{pW}$, which corresponds to $\mathcal{P}_{\text{THG}}=0.65\,\text{nW}$.}
\begin{tabularx}{1\columnwidth}{l@{\extracolsep{\fill}} c c } 
\hline\hline
Parameter & \multicolumn{1}{c}{Low Limit} & High Limit \\
\hline\hline
$\kappa_{3,\text{i+p}}/ 2\pi$ & \multicolumn{1}{c}{$2.13\,\text{THz}$} & $2.13\,\text{THz}$ \\
$\kappa_{3,\text{ex}}/ 2\pi$ & \multicolumn{1}{c}{$1.06\,\text{THz}$} & $6.15\,\text{GHz}$ \\
$\kappa_{3}/ 2\pi$ & \multicolumn{1}{c}{$4.25\,\text{THz}$} & $2.14\,\text{THz}$ \\
$P_{\text{circ}}^{\text{THG}}$ & \multicolumn{1}{c}{$2.8\,\text{nW}$} & $484\,\text{nW}$ \\
$\mathcal{F}_{\text{THG}}$ & \multicolumn{1}{c}{$13$} & $2358$ \\
\hline\hline
\label{table:Estimation_Green_Power}
\end{tabularx} 
\end{table}

\section{Thermo-optic effects}
\label{sec:Thermo-optic_effects}
Heating of the cavity, either internally by the pump laser or externally with a thermoelectric module, will result in red-shifting of the cavity mode due to thermo-optic effects related to refractive index modulation caused by changes in temperature, and due to thermal expansion of the cavity\,\cite{Sun2017OpticsExpress}. The dynamics of thermo-optic effects are governed by the heat equation\,\cite{Sun2017OpticsExpress,Hu2021PRL}
\begin{equation}
    \Delta\dot{\bar{T}} = - \Gamma_\text{T}\Delta\bar{T} + \eta_\text{T} |a_1|^2,
    \label{eq:thermal}
\end{equation}
where $\Delta\bar{T}$ is the absolute temperature change of the cavity, $\Gamma_\text{T}$ is thermal relaxation rate of the PCC, and $\eta_\text{T}$ relates the cavity field to the heat that it generates and will vary depending on the optical absorption properties of the diamond. Consequently, the coupled mode equation (\Cref{eq:CME_1}) describing the cavity dynamics must be modified with an additional thermal coupling term\,\cite{Sun2017OpticsExpress}, thus becoming
\begin{equation}
    \dot{a}_1 = \left(i\Delta_0 - \frac{\kappa_1}{2}\right)a_1 -ig_T\Delta\bar{T}a_1 + \sqrt{\kappa_{\text{1,ex}}}\;s_\text{1,in}\,,
    \label{eq:CME_thermal}
\end{equation}
where $g_T := \frac{d\omega_c}{d \Delta\bar{T}}|_{_{\Delta\bar{T} = 0}}$ is a thermal coupling coefficient derived from a series expansion of $\omega_c(\Delta\bar{T}) \approx \omega_0 + \frac{d\omega_c}{d\Delta\bar{T}}|_{_{\Delta\bar{T} = 0}}\Delta\bar{T}$. The timescale of the thermal effects acting on the cavity is on the order of $10^{-6}\,\text{s}$\,\cite{Lake2018}. We therefore assume a steady-state approximation with $\Delta\dot{\bar{T}}=0$. Using this assumption and solving for $\Delta \bar{T}$, we obtain $\Delta\bar{T}=\frac{\eta_\text{T}}{\Gamma_\text{T}}\left|a_1\right|^2$. Inserting this expression into \Cref{eq:CME_thermal} and solving for $a_1$ yields
\begin{equation}
    \label{Eq:thermal_a}
    a_1 = \frac{\sqrt{\kappa_{\text{1,ex}}}}{-i(\Delta_0 - c_{\text{T}}|a_1|^2) + \frac{\kappa_1}{2}}\:s_\text{1,in}\,.
\end{equation}
The intracavity photon number is given by $n_\text{cav} = |a_1|^2$. Hence,
\begin{equation}
    \label{eq:cubic_eq}
    n_\text{cav} = \frac{\kappa_{\text{1,ex}}}{\left(\Delta_0 - c_\text{T} n_\text{cav}\right)^2 + \left(\frac{\kappa_1}{2}\right)^2}\times\frac{\tilde{P}_{\text{in}}}{\hbar\omega_0}, 
\end{equation}
where
\begin{equation}
    c_\text{T} := \frac{g_\text{T}\eta_\text{T}}{\Gamma_\text{T}}
\end{equation}
is a thermo-optic coefficient. In general, cubic equations such as \Cref{eq:cubic_eq} have three complex solutions. However, here we only consider the real part of the solution that matches the experimental results. By inserting $N = \Re(n_\text{cav})$ into \Cref{eq:CME_thermal}, and utilising input-output formalism, we can express the normalised cavity transmission with thermo-optic effects on the cavity as 
\begin{equation}
    \label{eq:transmission_TO}
    T_{\text{TO}} = \left|e^{i\phi_1} - \frac{\kappa_{\text{1,ex}}}{-i(\Delta_0 - c_T N) + \frac{\kappa_1}{2}}\right|^2\,.
\end{equation}

\begin{figure}[t!]
    \centering
    \includegraphics[width=\columnwidth]{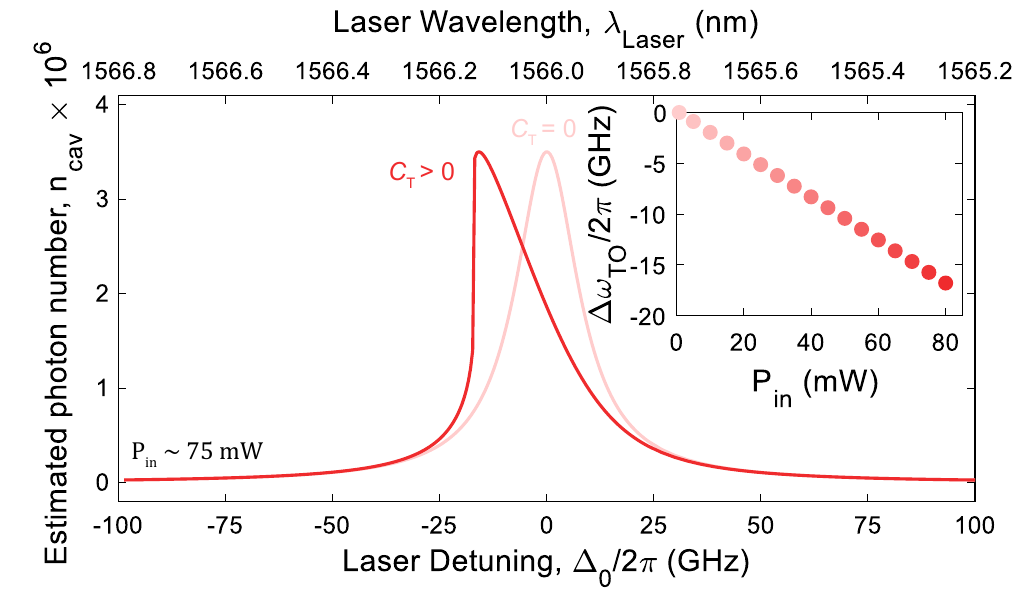}
    \caption{The solid red line shows the estimated intracavity photon number when $P_{\text{in}}\sim75\,\text{mW}$, calculated from \Cref{eq:cubic_eq}. For comparison, the pink line shows the estimated intracavity photon number in the absence of the thermo-optic effect, i.e. $c_\text{T} = 0$ for the same input power. The inset shows the input power dependence of the cavity red shift, $\Delta\omega_{\text{TO}}=-c_{\text{T}}n_{\text{cav}}$. 
    } 
    \label{fig:N_cav}
\end{figure}

We next estimate the IR intracavity photon number, $n_{\text{cav}}$, using \Cref{eq:cubic_eq}. In \Cref{fig:N_cav}, we plot the dependency of $n_{\text{cav}}$ with wavelength for the cavity mode at $1566\,\text{nm}$ and find that for input power $P_\text{in}\sim75\,\text{mW}$, the maximum IR input power applied in the THG power dependent measurement series, $n_\text{cav}\sim3.5\times10^{6}$. In this calculation, we have estimated the power at the fibre-taper cavity interface according to $\tilde{P}=\sqrt{\eta_{\text{fibre}}}P_{\text{in}}$, as outlined in \Cref{sec:setup}.
We note that although the cavity line shape is asymmetric due to photo-thermal effect, the increase in device temperature is insufficient to observe thermal bistable behaviour.

The thermo-optic effect can be used to red-shift the optical cavity.
The corresponding change in device temperature can be calculated from the relation\,\cite{Fontanella1977}
\begin{equation}
    \frac{1}{n}\left(\frac{\Delta n}{\Delta T}\right)=4.04\times10^{-6}\,\text{K}^{-1}\,.
\end{equation}
By extracting the initial negative frequency shift in Fig.\,3\,(b) of the main manuscript, we experimentally demonstrate the ability to red-shift the cavity resonance by $- 9.1\,(2)\,\text{GHz}$, which corresponds to an increase in device temperature of $\sim12\,\text{K}$.
Since the cavity detuning is proportional to $n_\text{cav}$ (\Cref{eq:transmission_TO}), increasing the input power would manifest in a larger red shift.  We therefore calculate the expected cavity red-shift, $\Delta\omega_{\text{TO}}=-c_{\text{T}}n_{\text{cav}}$, as a function of ${P}_{\text{in}}$, using \Cref{eq:cubic_eq}. The inset in \Cref{fig:N_cav} shows the result of these calculations. We find that for $P_{\text{in}}=100\,\text{mW}$, a cavity red-shift of $\Delta\omega_{\text{TO}}/2\pi=-19.1\,\text{GHz}$ is achievable.
This large red shift would correspond to a change in device temperature of $\sim25\,\text{K}$.
Note that optimising the device geometry to improve heat dissipation would allow for mitigation of the thermal red-shift for larger input powers. Similarly, a design with minimal heat dissipation would achieve even greater red shifts.

\begin{figure}[t!]
    \centering
\includegraphics[width=\columnwidth]{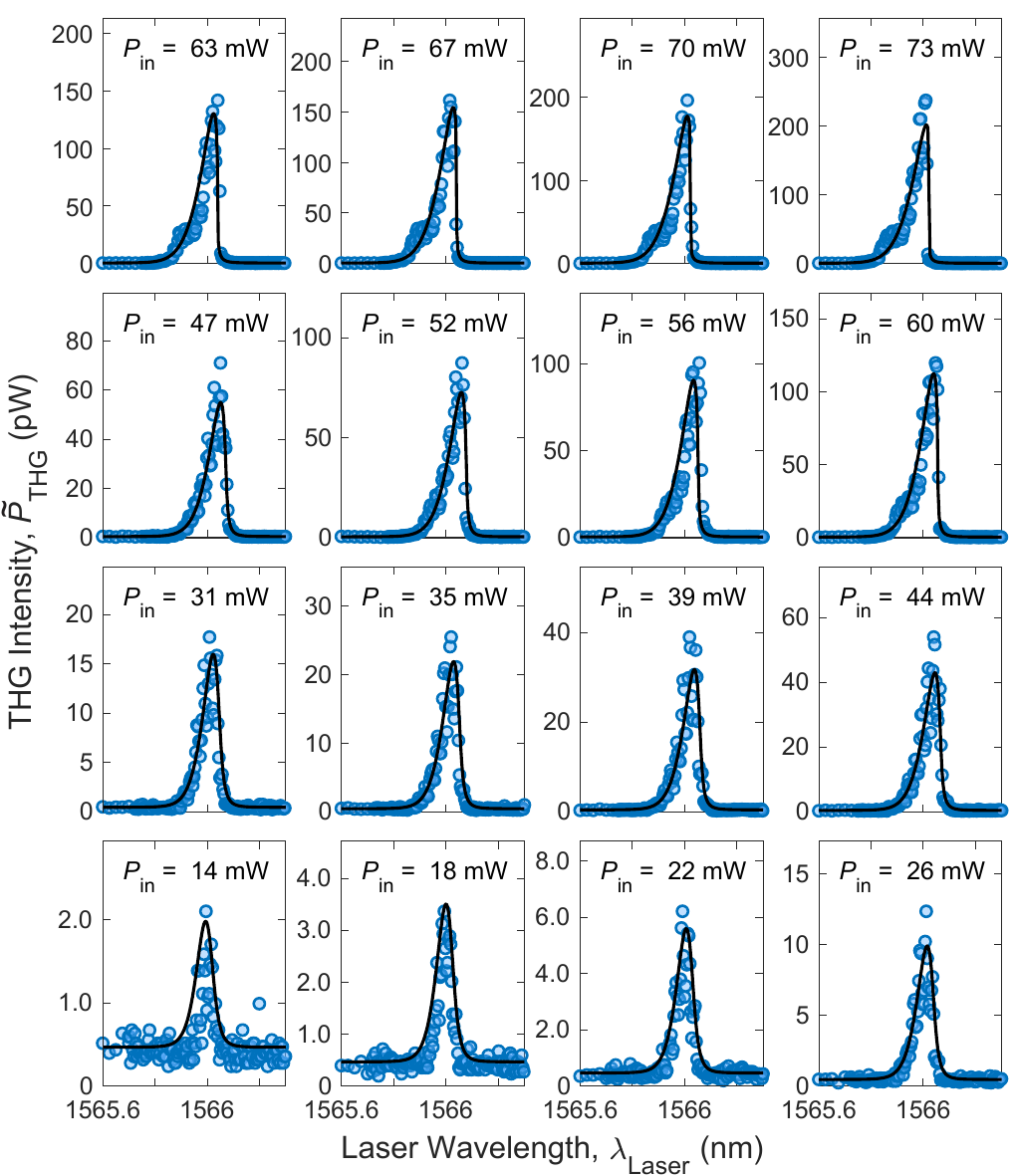}
    \caption{In blue, intensity of the THG signal plotted against the wavelength for different input powers. The black lines are a fit to the model (\Cref{eq:THG_power_thermal}). As discussed in the main text, for large input powers, the THG signal exhibit a shoulder on the red side of the resonance due to optomechanical self-sustained oscillations.
    }
    \label{fig:THG_all_fits}
\end{figure}

Finally, we derive an expression for the third-harmonic power exiting the fibre-taper in the presence of the thermo-optic effect using \Cref{Eq:thermal_a} and the input-output formalism
\begin{equation}
    \label{eq:THG_power_thermal}
    \mathcal{P}_{\text{THG}} = \frac{108}{\hbar^2}\times\mathcal{L}\times\left(\frac{\kappa_{\text{1,ex}}}{(\Delta_0-c_{\text{T}}N)^2 + (\frac{\kappa_1}{2})^2}\right)^3\;\tilde{P}_{\text{in}}^3\,.
\end{equation}
Note that this expression can be obtained more directly by substituting $\Delta_0$ in \Cref{eq:THG_power} with the thermo-optic detuning term $\Delta_{\text{TO}}=\Delta_0-c_{\text{T}}N$ derived in \Cref{Eq:thermal_a}. In \Cref{fig:THG_all_fits}, we use \Cref{eq:THG_power_thermal} to fit the data of each individual scan from the power dependent THG series shown in Fig.\,2\,(b) of the main manuscript. We find that the fit is in excellent concordance with the measured data for the range of investigated input powers.

\subsection{Optomechanical self-oscillations}
\begin{figure}[t!]
    \centering
    \includegraphics[width = \columnwidth]{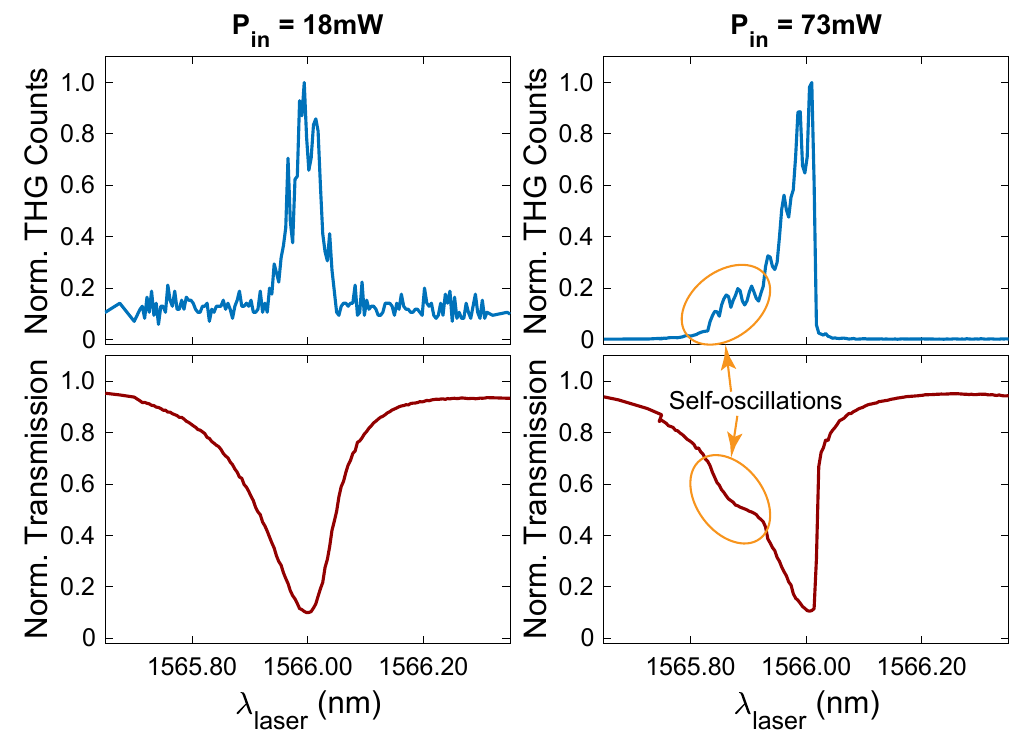}
    \caption{Large optical input powers result in mechanical self-oscillations,  manifested by the emergence of a shoulder on the blue side of the cavity resonance visible for both the pump cavity mode and the THG spectra. The threshold for self-oscillations depends on $n_{\text{cav}}$ and is therefore not visible for low input power.
    }
\label{fig:Self_Osc}
\end{figure}

The transmission and THG spectra in Fig.\,2\,(a) of the main manuscript show evidence of mechanical self-sustained oscillations. These oscillations occur for large optical input power, where optomechanical backaction exceeds the mechanical decay rate\,\cite{Aspelmeyer2014}. The self-oscillations cause the device to oscillate at the mechanical resonance frequency ($\sim10\,\text{GHz}$), which is manifested by the emergence of the shoulder seen on the blue side of the cavity resonance in Fig.\,2\,(a). For constant $\kappa_{1}$ and $\kappa_{1,\text{ex}}$, the threshold for self-oscillations depends on  $n_{\text{cav}}$ (see \Cref{fig:Self_Osc}). A thorough characterisation of the mechanical properties of the PCC is beyond the scope of this work.

\section{Photorefractive Effect}
\begin{figure}
    \centering
    \includegraphics[width = \columnwidth]{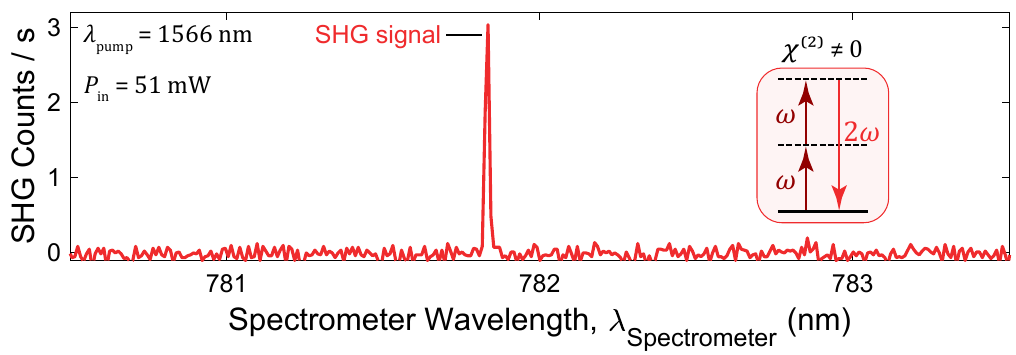}
    \caption{Second-harmonic generation from the diamond PCC, indicating a non-zero $\chi^{(2)}$. The inset shows the second-harmonic generation process, where two photons at frequency $\omega$ upconvert to one photon at $2\omega$.}
    \label{fig:SHG_single_sepctrum}
\end{figure}

The photorefractive effect constitutes a change of refractive index due to optically induced charge redistribution within a material\,\cite{Gunter1988,Johansen2003,Hou2024OpticsLett}. The photo-induced charge redistribution creates a space-charge electric field, $E_{\text{sp}}$, which modifies the refractive index via the electro-optic effect\,\cite{Liang2017Optica}. The observation of the linear electro-optic (Pockels) effect is unexpected in diamond due to its centrosymmetric crystal structure and consequently vanishing intrinsic $\chi^{(2)}$. However, breaking the crystal inversion symmetry results in  $\chi^{(2)}\neq0$, as observed in several recent studies\,\cite{Abulikemu2021, Abulikemu2022, Flagan2025}.
In \Cref{fig:SHG_single_sepctrum}, we show measurement of a second harmonic generation from the PCC, which demonstrates that it exhibits a non-zero $\chi^{(2)}$.

We now discuss the origin of $\chi^{(2)}\neq 0$. As mentioned above, the device studied in this work was fabricated from `optical grade' single-crystal diamond (Element Six). This material contains crystal defects, such as NV centres, substitutional nitrogen, N$_{\text{s}}$\,\cite{Ashfold2020}, and possibly additional nitrogen-vacancy complexes (N$_n$V)\,\cite{Ashfold2020,Bourgeois2022}.
Substitutional nitrogen impurities are the dominant defects at a concentration of $\sim 1\,\text{ppm}$, while the NV concentration is estimated to be $\sim 0.01\,\text{ppm}$\,\cite{Acosta2009}.
The NV centre exhibits two charge states: the negatively charged NV$^-$ and the neutral NV$^0$. For diamond with a high nitrogen concentration, NV$^-$ is the preferred charge state\,\cite{Manson2005}, which we confirm experimentally from previous photoluminescence studies performed on devices fabricated from similar diamond material\,\cite{Shandilya2021,Masuda2024,Shandilya2024arxiv,Flagan2025}.
Stabilisation of NV$^-$ requires the presence of electron donors in the lattice\,\cite{Santori2009} such as the aforementioned neutrally charged N$_{\text{s}}^{0}$\,\cite{Manson2005,Acosta2009,Fu2010,Haque2017,Luo2022APL}. Photoionisation of N$_{\text{s}}^{0}$ and subsequent electron capture by nearby NV$^0$ leads to the formation of NV$^{-}$--N$_{\text{s}}^{+}$ pairs\,\cite{Manson2018,Blakley2024}. These charged defects creates static electric fields, $E_{\text{DC}}$, which can couple to the intrinsic bulk $\chi^{(3)}$ to form an effective $\chi^{(2)}_{\text{eff}}=3\chi^{(3)}E_{\text{DC}}$\,\cite{Castellan2019,Flagan2025}.
In principle, symmetry breaking by surfaces\,\cite{Trojanek2010,Levy2011,Zhang2019} or the presence of defects\,\cite{Li2022CellReports,Abulikemu2021,Abulikemu2022} will also induce a $\chi^{(2)}>0$. 
However, in our earlier work, we have demonstrated that electric fields are the predominant contributors to $\chi^{(2)}_{\text{eff}}\neq 0$\,\cite{Flagan2025}. 
We therefore believe that the electric field-induced $\chi^{(2)}_{\text{eff}}$, albeit weak, is the source of the observed electro-optic effect.
It is also worth noting that electric field-induced electro-optic effects has been demonstrated in various silicon photonic platforms\,\cite{Zhu2012Optics, Castellan2019Frontiers, Chakraborty2020, Friedman2021, Zabelich2022, Peltier2024}.

\begin{figure}[t!]
    \centering
    \includegraphics[width=\columnwidth]{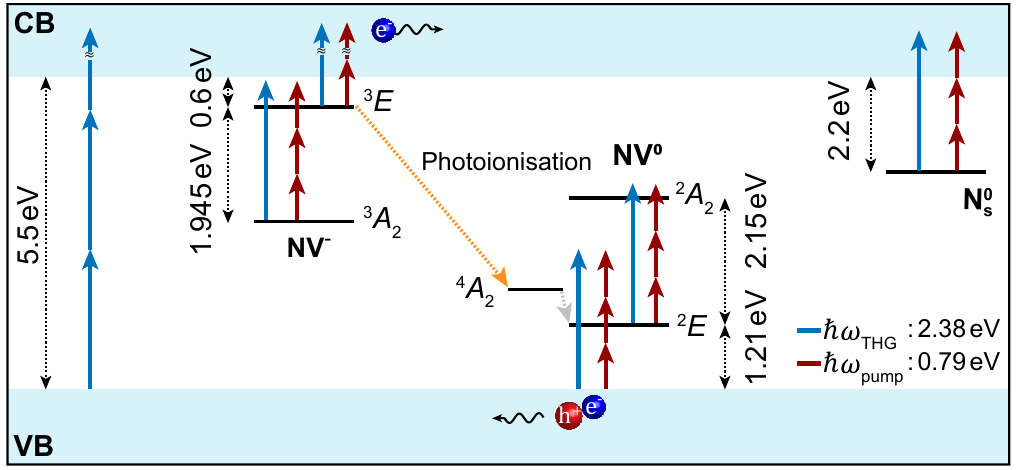}
    \caption{Relative position of the various defect levels within the bandgap of the diamond. Third-harmonic-generated photons are sufficiently energetic to optically excite both charge states of the NV centre and N$_{\text{s}}$. Photoionisation from NV$^-$ to NV$^0$ liberates an electron to the conduction band and occurs from $^3E$ excited state of NV$^-$ by the absorption of either one THG photon or two IR photons, which transfers the system to the $^4A_2$ quartet state of NV$^0$. Nonradiative relaxation brings the system from $^4A_2$ to the $^2E$ ground state of NV$^0$.
    Recombination from NV$^0$ to NV$^-$ requires optical excitation to the $^2A_2$ excited state of NV$^0$ followed by electron capture from the valence band, which can be achieved by the absorption of THG photons. The orange dotted arrow indicates photoionisation from NV$^-$ to NV$^0$, while the grey dotted arrow indicates the nonradiative decay from $^4A_2$ to $^2E$.
    } 
    \label{fig:Energy_Levels}
\end{figure}

We next discuss possible mechanisms for photo-induced charge redistribution and the formation of the space-charge electric field, $E_{\text{sp}}$. Figure\,\Cref{fig:Energy_Levels} shows the relevant energy levels for the dominant crystal defects and their possible photoionisation processes. Optical excitation and photoionisation of these defects, either by multi-photon absorption of IR pump photons ($\hbar\omega_{1566\,\text{nm}}=0.79\,\text{eV}$), the absorption of the internally generated third-harmonic photons ($\hbar\omega_{522\,\text{nm}}=2.38\,\text{eV}$), or the absorption of a green photon followed by absorption of two IR photons, liberate electrons to the conduction band.
The free electrons diffuse to regions of low intensity under the influence of the strong IR field, which facilitates the formation of the space-charge electric field, $E_{\text{sp}}$\,\cite{Johansen2003}. 
We note that due to the highly localised nature of the IR field (\Cref{fig:Comsol}\,(c)), the cavity mode is only affected by changes to the refractive index in the centre region of the device.

As illustrated in Fig.\,\ref{fig:Energy_Levels}, due to the large bandgap of diamond ($\sim5.5\,\text{eV}$\,\cite{Ashfold2020}), direct photoionisation from the valence to the conduction band would require absorption of three THG photons and is considered unlikely. We next consider N$_{\text{s}}^{0}$, whose photoionisation threshold of $2.2\,\text{eV}$\,\cite{Rosa1999,Bourgeois2022,Orphal-Kobin2023} can be bridged by the absorption of a single THG photon.
The comparatively large concentration of N$_{\text{s}}$ suggests that the photoionisation process N$_{\text{s}}^{0}\rightarrow$N$_{\text{s}}^{+}+e^{-}$ is the dominant contributor to the liberation of electrons leading to the charge redistribution\,\cite{Goldblatt2026}.
Photoionisation of NV$^-$ to NV$^0$ constitutes another candidate for charge redistribution. The threshold for direct photoionisation from the ${^3A_2}$ ground-state of NV$^-$ to the ${^2E}$ ground-state of NV$^0$ has been measured experimentally to be $2.65\,\text{eV}$\,\cite{Aslam2013,Bourgeois2017}, which would require absorption of multiple THG photons and is therefore expected to contribute only weakly to the charge redistribution.
Alternatively, photoionisation can take place from the ${^3E}$ excited state, which lies $1.95\,\text{eV}$ above the ${^3A_2}$ ground state\,\cite{Doherty2011}. A single THG photon is sufficiently energetic to populate the ${^3E}$ excited state.
There are then two possible photoionisation processes from the ${^3E}$ state: ${^3E}\rightarrow {^2A_2}$ and ${^3E}\rightarrow {^4A_2}$\,\cite{Shandilya2024arxiv}. The energy threshold for the photoionisation process ${^3E}\rightarrow{^2A_2}$ is $2.86\,\text{eV}$\,\cite{Shandilya2024arxiv}.
Again, as this would require multi-photon absorption, we expect limited contribution from this process to the charge redistribution.
This leaves the process ${^3E}\rightarrow {^4A_2}$, from which non-radiative decay relaxes the system to the ${^2E}$ ground state of NV$^0$\,\cite{Shandilya2024arxiv}. Although the exact energy difference between the states ${^4A_2}$ and ${^2E}$ is not known experimentally\,\cite{Razinkovas2021Photoionization,Thiering2024}, a recent experimental study restricted the photoionisation threshold for the process ${^3E}\rightarrow {^4A_2}$ to $>1.18\,\text{eV}$\,\cite{Shandilya2024arxiv}.
The photoionisation process ${^3E}\rightarrow {^4A_2}$ is therefore achievable via the absorption of one THG or two IR pump photons\,\cite{Shandilya2024arxiv}. 
For the large IR field intensity achieved within the PCC in this experiment, we expect the two-photon IR photoionisation process from $^3E$ to be a dominant contributor to the liberation of electrons\,\cite{Shandilya2024arxiv}. 
This two-photon process is localised to the centre region of the device, where the IR field strongly confined.
We note that the effect of the strong IR field on the photophysics of N$_{\text{s}}$ remains unknown\,\cite{Shandilya2024arxiv}. 
Furthermore, it is crucial to emphasise that in the absence of the THG light, no charge dynamics will occur and the system will remain in the ground state of NV$^{-}$.
The importance of green light was further demonstrated in Fig.\,5 of the main manuscript, where the photorefractive blue-shift was found to accelerate when the device was simultaneously exposed to IR and green laser fields.
The internal generation of green light is therefore of fundamental importance to the observation of the photorefractive effect. 

Finally, recombination from NV$^{0}$ to NV$^{-}$ is possible by optical excitation of NV$^{0}$ followed by subsequent electron capture from the valence band. The THG photons are sufficiently energetic to allow for this recombination process. 
For future experimental studies, measurements of the photorefractive effect can be performed in conjunction with photoluminescence measurements of the NV centre to investigate the resultant charge-state dynamics\,\cite{AudeCraik2020,Blakley2024,Shandilya2024arxiv}. Alternatively, additional investigation into the local charge environment and charge-transfer between defects\,\cite{KuateDefo2023} can be performed by utilising optically detected magnetic resonance schemes\,\cite{Broadway2018,Mittiga2018,Yuan2020PhysRevResearch,Block2021}, photoconductivity measurements\,\cite{Bourgeois2022,Todenhagen2025}, electrometers based on point defects\,\cite{Pieplow2025,Goldblatt2026}, or by measuring the spectral diffusion dynamics of the NV$^{-}$ zero-phonon line at cryogenic temperatures\,\cite{Orphal-Kobin2023}. 

In summary, there are two mechanisms contributing to the observed electro-optic effect. First, charged defects induce a static electric field, $E_\text{DC}$, which couples to the intrinsic bulk $\chi^{(3)}$ to form an effective $\chi^{(2)}_{\text{eff}}=3\chi^{(3)}E_{\text{DC}}$\,\cite{Flagan2025}. This non-zero $\chi^{(2)}$ enables the observation of the electro-optic effect. Next, photoionisation of crystal defects inside regions of high intensity, predominately N$^{0}_{\text{s}}$, liberate electrons into the conduction band\,\cite{Johansen2003}. Diffusion of these free electrons into regions of low intensity leads to the build-up of a space-charge field, $E_\text{sp}$. The observed photorefractive effect is the consequence of the electro-optic interaction between $\chi^{(2)}_{\text{eff}}$ and $E_\text{sp}$.

\subsection{Modelling the Photorefractive Resonance Shift}
The PCC presented in this work is an ideal platform to study photorefraction in diamond, owing to its sensitivity to changes in the refractive index. 
In the PCC, photorefraction manifests as a blue shift of the cavity resonance frequency, driven using a strong resonant pump laser.
The pump laser, through photoionisation and charge diffusion, generates a space-charge field, $E_{\text{sp}}$, whose time evolution can be described by\,\cite{Sun2017OpticsExpress}
\begin{equation}
    \dot{E}_\text{sp}= - \Gamma_{\text{E}}E_{\text{sp}} +\eta_{\text{E}}|a_1|^2\,.
    \label{eq:E_sp_dot}
\end{equation}
The equation for the space-charge field looks similar to the thermo-optic equation (\Cref{eq:thermal}). Here, $\Gamma_{\text{E}}$ is the space-charge field relaxation rate, describing how fast the space-charge dissipates in the cavity, and $\eta_{\text{E}}$ is a coefficient that described the rate at which the space-charge field is generated per photon\,\cite{Sun2017OpticsExpress}. 
We assume that the only time-dependent contributions come from the space-charge field. Assuming no initial space-charge field, $E_\text{sp}(0) = 0$, the solution to \Cref{eq:E_sp_dot} is
\begin{equation}
    \label{eq:ODE_solution_PR}
    E_{\text{sp}}(t) = \frac{\eta_{\text{E}}}{\Gamma_{\text{E}}}\left(1 - e^{-\Gamma_{\text{E}}t}\right)\left|a_1\right|^2\,.
\end{equation}

As before, the coupled mode equation for the cavity field (\Cref{eq:CME_1}) needs to be modified to include the photorefractive coupling term $-ig_{\text{E}}E_{\text{sp}}$.
We use a Taylor expansion of the cavity resonance, $\omega_\text{c}(E_\text{sp}) \approx \omega_0 + \frac{d\omega_\text{c}}{dE_\text{sp}}|_{_{E_\text{sp}=0}}E_\text{sp}$, and define the coefficient $g_\text{E}:= \frac{d\omega_\text{c}}{dE_\text{sp}}|_{_{E_\text{sp} = 0}}$.
The term $g_\text{E}$ is proportional to the electro-optic coefficients\,\cite{Sun2017OpticsExpress}. 
The modified equation of motion becomes
\begin{equation}
    \label{eq:PR}
    \dot{a}_1 = \left(i\Delta_0 - \frac{\kappa_1}{2}\right)a_1 - ig_{\text{E}}E_{\text{sp}}a_1 +\sqrt{\kappa_{\text{1,ex}}}\:s_\text{1,in}\,.
\end{equation}
As before, we assume a steady-state solution for the optical field amplitude, i.e. $\dot{a}_1=0$. Solving for $a_1$ and utilizing input-output formalism, we derive an expression for the cavity transmission
\begin{equation}
    \label{eq:transmission_PR}
    T_\text{PR} = \left|e^{i\phi_1} - \frac{\kappa_{\text{1,ex}}}{-i\left(\Delta_0 - c_{E}\left(1 - e^{-\Gamma_{\text{E}}t}\right)\right) + \frac{\kappa_1}{2}}\right|^2\,,
\end{equation}
where 
\begin{equation}
    c_{\text{E}}=\frac{g_{\text{E}}\eta_{\text{E}}}{\Gamma_{\text{E}}}|a_1|^2
\end{equation}
is a photorefractive coefficient.

\section{Cavity mode coupling to thermal and photorefractive effects}
\label{sec:Mode_Coupling_Thermal_PR_Effect}
For a large laser input power ($P_{\text{in}}\sim70\,\text{mW}$), we observed that the PCC exhibits simultaneous photorefractive and thermo-optic effects, resulting in a net blue shift of the cavity resonance. To describe the resultant dynamics from these two independent processes, we combine  \Cref{eq:CME_1,eq:thermal,eq:E_sp_dot}.
The cavity resonance frequency depends on both the absolute change of the cavity temperature and the cavity space-charge field. We implement a Taylor expansion with $\Delta\bar{T}\ll1$ and $E_\text{s} \ll 1$, which gives $\omega_\text{c}(\Delta\bar{T},E_\text{sp}) \approx \omega_0 + \frac{\partial \omega_\text{c}}{\partial \Delta\bar{T}}|_{_{\Delta\bar{T} = 0}} \Delta\bar{T} + \frac{\partial \omega_\text{c}}{\partial E_\text{sp}}|_{_{E_\text{sp} = 0}}E_\text{sp}$. We find
\begin{align}
    \label{eq:CME_TO_PR}
    \dot{a}_1 &= \left(i\Delta_0 - \frac{\kappa_1}{2}\right)a_1 - ig_{\text{T}}\Delta \bar{T}a_1 - ig_{\text{E}}E_{\text{sp}}a_1 + \sqrt{\kappa_{\text{ex},1}}\:s_\text{1,in}\,, \\
    \label{eq:TO}
     \Delta\dot{\bar{T}} &= -\Gamma_\text{T}\Delta \bar{T} + \eta_\text{T}\left|a_1\right|^2\,,\\
     \label{eq:PR_spaceCharge}
    \dot{E}_\text{sp} &= -\Gamma_\text{E}E_\text{sp} + \eta_\text{E}\left|\tilde{a}_1\right|^2\,.
\end{align}
As we discussed in the previous section, we assume that the only time-dependent contribution in \Cref{eq:PR_spaceCharge} is the evolution of the space-charge field $E_{\text{sp}}$. We assume a steady-state approximation for the other contributions, i.e. $\dot{a}_1 = 0$ and $\Delta \dot{\bar{T}} = 0$. Additionally, to be consistent with our experimental procedure, the laser pump wavelength is set to the cavity resonance, i.e. $\Delta=\Delta_0  - g_\text{T}\Delta \bar{T} - g_\text{E}E_\text{sp} = 0$. Therefore, we express the  number of photons on resonance as 
\begin{equation}
    \left|\tilde{a}_1\right|^2 =\left|{a}_1(\Delta=0)\right|^2= \frac{4\kappa_\text{1,ex}\left|s_\text{1,in}\right|^2}{\kappa_1^2}\,
\end{equation} 
in \Cref{eq:PR_spaceCharge}. As before, we use input-output formalism to derive an expression for the normalised cavity transmission. To that end, we insert $\Delta \bar{T}=\frac{\eta_{\text{T}}}{\Gamma_{\text{T}}}\left|a_1\right|^2$ and \Cref{eq:ODE_solution_PR} into \Cref{eq:CME_TO_PR} and solve for $a_1$. We find the fibre-taper transmission to be
\begin{equation}
    T = \left|e^{i\phi_1} - \frac{\kappa_{1,\text{ex}}}{-i\left(\Delta_0 - c_{\text{T}}N - c_{\text{E}}(1 - e^{-\Gamma_{\text{E}}t})\right) + \frac{\kappa_1}{2}}\right|^2\,.
    \label{eq:Tran_Thermal_Photor_theo}
\end{equation}

In the experimental data shown in Fig.\,3\,(b) of the main manuscript, we observe that the photorefractive blue-shifting occurred at two different time scales. We therefore fit the data using a double exponential model with both a fast and a slow time constant. We speculate that the two time constants stem from charges associated with the surface and the bulk, as observed for lithium niobate\,\cite{Jiang2017OptLett,Sun2017OpticsExpress,Ren2025}. Therefore, when fitting the experimentally observed photorefractive shift, we modify \Cref{eq:Tran_Thermal_Photor_theo} and use 
$\Delta\omega_\text{PR}(t) = C_\text{f}^{\text{PR}}(1-e^{-\Gamma_\text{f} t}) + C_\text{s}^{\text{PR}}(1-e^{-\Gamma_\text{s} t})$, where $C_\text{f,(s)}^{\text{PR}}$ is a fast (slow) photorefractive coefficient and $\Gamma_\text{f,(s)}$ is a fast (slow) space-charge relaxation rate. Thus, by considering the two different timescales, the fibre-taper transmission in the presence of the thermo-optic effect and photorefraction is given by
\begin{equation}
    \label{eq:transmission_PR_actual}
    T_{\text{PR}}(t) = \left|e^{i\phi_1} - \frac{\kappa_\text{1,ex}}{-i\left(\Delta_0 - c_\text{T}N - \Delta\omega_\text{PR}(t)\right) + \frac{\kappa_1}{2}}\right|^2\,.
\end{equation}

\begin{figure}[t!]
    \centering
    \includegraphics[width=\columnwidth]{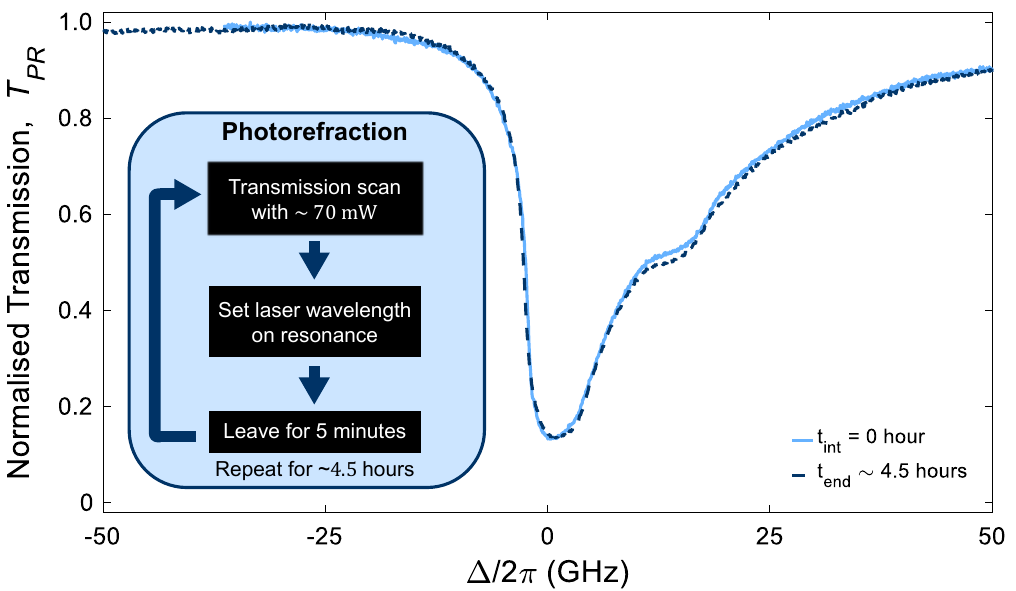}
    \caption{Comparison of the cavity line shape between the first ($t_{\text{int}}=0\,\text{hours}$) and last ($t_{\text{end}}\sim4.5\,\text{hours}$) laser transmission scans for the photorefractive data shown in Fig.\,3\,(a) of the main text. The detuning, $\Delta$, is artificially set for both fibre-taper transmission curves so that they overlap. There is no change to the fibre-taper transmission spectrum for the duration of the experiment -- photorefraction does not alter the cavity $Q$-factor.}
    \label{fig:PRE_transmission_overlap}
\end{figure}

\subsection{Experimentally Measuring the Photorefractive Resonance Shift}
We use the procedure illustrated in the inset of \Cref{fig:PRE_transmission_overlap} to measure the simultaneous thermal and photorefractive effects. In this experiment, the laser power is kept constant at $P^{\textrm{high}}_{\text{in}}\sim70\,\text{mW}$. First, a laser transmission scan is performed to locate the cavity resonance. Next, the laser is parked on the cavity resonance for five minutes. Then, another laser transmission scan is used to locate the new cavity resonance frequency. This procedure is repeated 50 times. 

In \Cref{fig:PRE_transmission_overlap}, we compare the cavity line shape of the first and last laser transmission scans by superimposing one on top of the other. We find that the cavity line shape remains unchanged for the duration of the measurement---there is no change to the $Q$-factor or the resonance contrast. We therefore conclude that the observed blue-shifting of the cavity resonance cannot be explained by changes to the coupling between the fibre-taper and the cavity. From the constant $Q$-factor, we further conclude that photorefraction does not introduce additional loss mechanisms, as previously observed in Z-cut lithium niobate microrings\,\cite{Ren2025}.

\section{Relaxation of photorefractive effects}
\label{sec:Relaxation}
\begin{figure}[t!]
    \centering
    \includegraphics[width=\columnwidth]{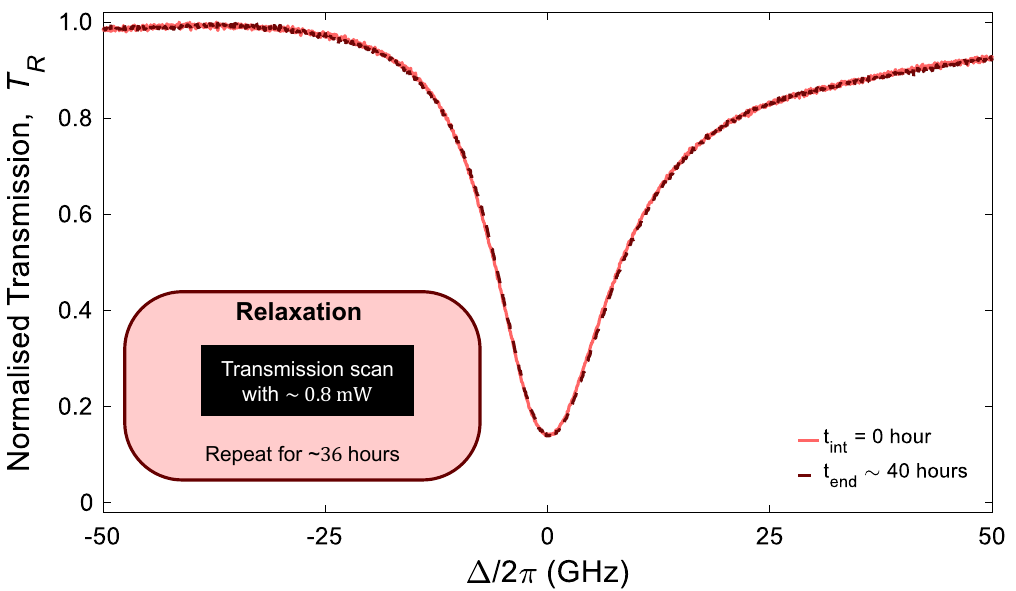}
    \caption{Comparison and overlap of the cavity line shape between the first ($t_{\text{int}}=0\,\text{hours}$) and last ($t_{\text{end}}\sim40\,\text{hours}$) laser transmission scans for the photorefractive relaxation data shown in Fig.\,3\,(c) of the main text. There is no change to the fibre-taper transmission spectrum for the duration of the experiment.
    }
\label{fig:Relax_transmission_overlap}
\end{figure}

After exploring the photorefractive blue-shifting of the cavity resonance for a large input power, we measure the relaxation of the photorefractive effect. To this end, following the inset of \Cref{fig:Relax_transmission_overlap}\,(b), we continuously scan the laser across the cavity for low input power ($P_\text{in}\sim 0.8\,\text{mW}$ and record the frequency of the cavity resonance. The resultant relaxation is shown in Fig.\,3\,(b) of the main manuscript. In \Cref{fig:Relax_transmission_overlap}, we superimpose the first and last laser transmission scan on top of each other. As for the photorefractive data, we observe no change to the $Q$-factor or the resonance contrast---red-shifting of the cavity resonance cannot be explained by movement of the fibre-taper.

We next derive a model explaining the cavity transmission in the presence of photorefractive relaxation. First, due to the low input power used to probe the relaxation, no thermo-optic effect is present. Second, we assume that the initial condition for the space-charge field, $E_\text{sp}^{\text{R}}$, for the relaxation process in \Cref{eq:E_sp_dot} is $E_\text{sp}^{\text{R}}(t=0)=E_\text{sp}(t\rightarrow\infty)$, thus $E_\text{sp}^{\text{R}}(t=0) = \frac{\eta_\text{E}}{\Gamma_\text{E}}\left|\tilde{a}_{1}\right|^2$. Therefore, the photorefractive relaxation process of the space-charge field is given by
\begin{equation}
    E_\text{sp}^{\text{R}} = \frac{\eta_\text{E}}{\Gamma_\text{E}}\left|\tilde{a}_{1}\right|^2 e^{-\Gamma_\text{E}t}\,.
\end{equation}
Following the procedure described in Sec.\,\ref{sec:Mode_Coupling_Thermal_PR_Effect}, the cavity transmission during the relaxation is given by
\begin{equation}
    \label{eq:transmission_relax}
    T = \left|e^{i\phi_1} - \frac{\kappa_\text{1,ex}}{-i\left(\Delta_0 - 
 c_\text{E}e^{-\Gamma_\text{E}t}\right) + \frac{\kappa_1}{2}}\right|^2\,.
\end{equation}
Similarly to the photorefractive blue-shifting described above, we observed that the relaxation process occurred at two different time scales. Therefore, we express the photorefractive relaxation term as 
$\Delta\omega_\text{R}(t) = \tilde{C}_\text{f}^{\text{R}}e^{-\gamma_\text{f}t} + \tilde{C}_\text{s}^{\text{R}}e^{-\gamma_\text{s}t}$, in which $\tilde{C}_\text{f,(s)}^{\text{R}}$ and $\gamma_\text{f,(s)}$ are the fast (slow) photorefractive relaxation coefficient and relaxation rate, respectively. Consequently, we modify \Cref{eq:transmission_relax} and derive a new cavity transmission function that considers photorefractive relaxation: 
\begin{equation}
    \label{eq:transmission_relax_actual}
    T_{\text{R}}(t) = \left|e^{i\phi_1} - \frac{\kappa_\text{1,ex}}{-i\left(\Delta_0  - \Delta\omega_\text{R}(t)\right) + \frac{\kappa_1}{2}}\right|^2\,.
\end{equation} 

\section{Reproducibility}
\begin{figure}[t!]
    \centering
    \includegraphics[width=\columnwidth]{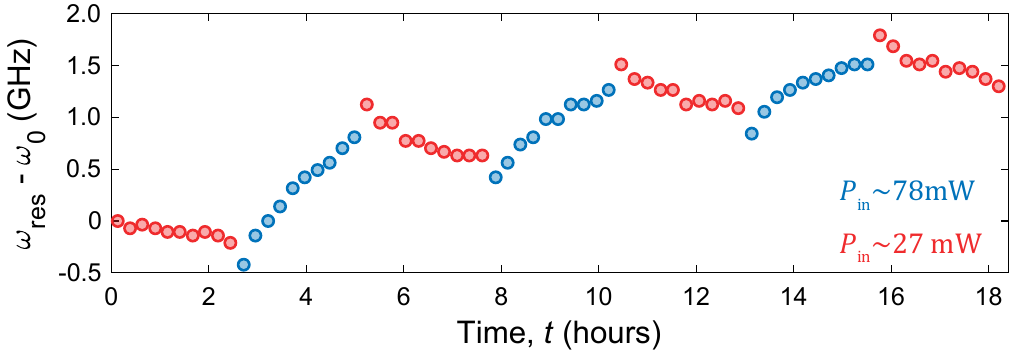}
    \caption{Demonstration of \textit{in situ} deterministic and reproducible blue-tuning via the photorefractive effect. 
    An alternating sequence of ten successive laser transmission scans performed for $P_{\text{in}}^{\text{low}}\sim27\,\text{mW}$ (red points) followed by ten laser scans for $P_{\text{in}}^{\text{high}}\sim78\,\text{mW}$ (blue points). The cavity resonance frequency blue-shifts when exposed to high-power IR, whereas a red-shift is observed for exposure to low-power IR, and this cycle is reducible over the measurement duration of 18 hours.
    }
\label{fig:Repeated_Measurements}
\end{figure}

We next turn to verify the reproducibility of the observed resonance blue-shifting. To this end, we devise a measurement scheme where the IR laser is step-wise tuned across the cavity---each wavelength step lasts for a duration of one second. We start by performing 10 successive scans for $P_{\text{in}}^{\text{low}}\sim27\,\text{mW}$ followed by 10 successive scans for $P_{\text{in}}^{\text{high}}\sim78\,\text{mW}$. We repeat this procedure three times, and the results are shown in \Cref{fig:Repeated_Measurements}.
A blue-shift of the cavity resonance is observed whenever the high-power laser is injected into the cavity, while the cavity resonance frequency red-shifts for low laser power.
Although an overall blue-shift is observed over the duration of the entire measurement sequence, \Cref{fig:Repeated_Measurements} clearly demonstrates that \textit{in situ} deterministic blue-tuning of the cavity resonance frequency is achievable by careful toggling of the laser power.
From this demonstration, we conclude that the photorefractive tuning mechanism is reproducible.
Finally, we note that the experimental procedure used in this measurement differs from the protocol used for Fig.\,(3) of the main manuscript, where the high-power laser was parked on resonance for 5 minutes (see Sections\,\ref{sec:Mode_Coupling_Thermal_PR_Effect} and \ref{sec:Relaxation}). Therefore, differences in shifting rates can be attributed to the difference in experimental protocol and in particular the laser exposure time on resonance.

\section{Summary of Fit Parameters}
In Table.\,\ref{table:fit_summary}, we summarise all the parameters extracted from the fits to the experimental data shown in the main text.

\begin{table}[h!]  
\centering 
\caption{Summary of fit parameters} 
\begin{tabularx}{1\columnwidth}{l @{\extracolsep{\fill}} c c c c} 
\hline\hline   
Parameter  & Value & Unit & Eq.  &
\\ 
\hline\hline  
$\tilde{\eta}_\text{THG}$ & $6\times10^{-7}$ & $\text{W}^{-2}$  &  \ref{eq:Estimated_THG_eff} & \\
$c_\text{T}/2\pi$ & $-4.53\,(7)$ & $\text{kHz}$ &  \ref{eq:transmission_TO}, \ref{eq:THG_power_thermal} & \\
$\mathcal{L} = \frac{\kappa_\text{3,ex}}{\kappa_3^2}|\beta_\text{THG}|^2$ & $2.9\,(3)\cdot 10^{-44}$ & $\text{s}$   & \ref{eq:THG_power_thermal}\\
$c_\text{T}/2\pi$ & $-2.9\,(2)$ & $\text{kHz}$ & \ref{eq:transmission_PR_actual} & \\
$C_\text{f}^{\text{PR}}/2\pi$ & $8\,(1)$ & $\text{GHz}$ & \ref{eq:transmission_PR_actual}\\
$\Gamma_\text{f}$ & $5\,(1)$ & $\text{hour}^{-1}$ & \ref{eq:transmission_PR_actual}\\
$C_\text{s}^{\text{PR}}/2\pi$ & $18.2\,(6)$ & $\text{GHz}$ & \ref{eq:transmission_PR_actual}\\
$\Gamma_\text{s}$ & $0.48\,(7)$ & $\text{hour}^{-1}$ & \ref{eq:transmission_PR_actual}\\
$\tilde{C}_\text{f}^{\text{R}}/2\pi$ & $5.5\,(2)$ & $\text{GHz}$ & \ref{eq:transmission_relax_actual}\\
$\gamma_\text{f}$ & $4.3\,(4)$ & $\text{hour}^{-1}$ & \ref{eq:transmission_relax_actual}\\
$\tilde{C}_\text{s}^{\text{R}}/2\pi$ & $4.1\,(1)$ & $\text{GHz}$ & \ref{eq:transmission_relax_actual}\\
$\gamma_\text{s}$ & $0.17\,(7)$ & $\text{hour}^{-1}$ & \ref{eq:transmission_relax_actual}\\
\hline 
\hline
\end{tabularx} 
\label{table:fit_summary}
\end{table} 


%